\RequirePackage{lineno}
\documentclass[twocolumn,aps,prd,floatfix,superscriptaddress]{revtex4}
\usepackage{latexsym, amsmath, amssymb,amsthm,amsopn,amsfonts, graphicx, epstopdf, multirow}
\usepackage{natbib}
\usepackage{graphicx}
\usepackage{dcolumn}
\usepackage{bm}
\usepackage{xcolor}
\usepackage{orcidlink}

\def\bicep{BICEP}

\def\bicepone{{\sc BICEP1}}
\def\biceptwo{{\sc BICEP2}}
\def\bicepthree{{\sc BICEP3}}

\def\keck{{\it Keck}}
\def\keckarray{{\it Keck Array}}

\def\planck{{\it Planck}} 

\def\wmap{WMAP}

\def\deg{^\circ}
\def\emode{$E$-mode}
\def\bmode{$B$-mode}

\def\lcdm{$\Lambda$CDM}

\begin{document}

\preprint{APS/123-QED}

\title{BICEP/\textit{Keck} XX: Component-separated maps of polarized CMB and thermal dust emission using \textit{Planck} and BICEP/\textit{Keck} Observations through the 2018 Observing Season}

\author{BICEP/\textit{Keck} Collaboration:~P.~A.~R.~Ade}
\affiliation{School of Physics and Astronomy, Cardiff University, Cardiff, CF24 3AA, United Kingdom}

\author{Z.~Ahmed\orcidlink{0000-0002-9957-448X}}
\affiliation{Kavli Institute for Particle Astrophysics and Cosmology, Stanford University, Stanford, CA 94305, USA}
\affiliation{SLAC National Accelerator Laboratory, Menlo Park, CA 94025, USA}

\author{M.~Amiri\orcidlink{0000-0001-6523-9029}}
\affiliation{Department of Physics and Astronomy, University of British Columbia, Vancouver, British Columbia, V6T 1Z1, Canada}

\author{D.~Barkats\orcidlink{0000-0002-8971-1954}}
\affiliation{Center for Astrophysics, Harvard \& Smithsonian, Cambridge, MA 02138, USA}

\author{R.~Basu Thakur\orcidlink{0000-0002-3351-3078}}
\affiliation{Department of Physics, California Institute of Technology, 1200 E. California Boulevard, Pasadena, CA 91125, USA}

\author{C.~A.~Bischoff\orcidlink{0000-0001-9185-6514}}
\affiliation{Department of Physics, University of Cincinnati, Cincinnati, OH 45221, USA}

\author{D.~Beck\orcidlink{0000-0003-0848-2756}}
\email{dobeck@stanford.edu}
\affiliation{Department of Physics, Stanford University, Stanford, CA 94305, USA}

\author{J.~J.~Bock}
\affiliation{Department of Physics, California Institute of Technology, 1200 E. California Boulevard, Pasadena, CA 91125, USA}
\affiliation{Jet Propulsion Laboratory, California Institute of Technology, Pasadena, CA 91109, USA}

\author{H.~Boenish}
\affiliation{Center for Astrophysics, Harvard \& Smithsonian, Cambridge, MA 02138, USA}

\author{V.~Buza}
\affiliation{Kavli Institute for Cosmological Physics, University of Chicago, Chicago, IL 60637, USA}

\author{B.~Cantrall\orcidlink{0000-0003-4541-7080}}
\affiliation{Department of Physics, Stanford University, Stanford, CA 94305, USA}
\affiliation{Kavli Institute for Particle Astrophysics and Cosmology, Stanford University, Stanford, CA 94305, USA}

\author{J.~R.~Cheshire IV\orcidlink{0000-0002-1630-7854}}
\affiliation{Department of Physics, California Institute of Technology, 1200 E. California Boulevard, Pasadena, CA 91125, USA}

\author{J.~Connors}
\affiliation{National Institute of Standards and Technology, Boulder, CO 80305, USA}

\author{J.~Cornelison\orcidlink{0000-0002-2088-7345}}
\affiliation{Argonne National Laboratory, High Energy Physics Division, Lemont, IL 60439, USA}

\author{M.~Crumrine}
\affiliation{School of Physics and Astronomy, University of Minnesota, Minneapolis, MN 55455, USA}

\author{A.~J.~Cukierman}
\affiliation{Department of Physics, California Institute of Technology, 1200 E. California Boulevard, Pasadena, CA 91125, USA}

\author{E.~Denison}
\affiliation{National Institute of Standards and Technology, Boulder, CO 80305, USA}

\author{L.~Duband}
\affiliation{Service des Basses Temperatures, Commissariat a l’Energie Atomique, 38054 Grenoble, France}

\author{M.~Echter}%
\affiliation{Center for Astrophysics, Harvard \& Smithsonian, Cambridge, MA 02138, USA}

\author{M.~Eiben}
\affiliation{Center for Astrophysics, Harvard \& Smithsonian, Cambridge, MA 02138, USA}

\author{B.~D.~Elwood\orcidlink{0000-0003-4117-6822}}
\affiliation{Department of Physics, Harvard University, Cambridge, MA 02138, USA}
\affiliation{Center for Astrophysics, Harvard \& Smithsonian, Cambridge, MA 02138, USA}

\author{S.~Fatigoni\orcidlink{0000-0002-3790-7314}}
\affiliation{Department of Physics, California Institute of Technology, 1200 E. California Boulevard, Pasadena, CA 91125, USA}

\author{J.~P.~Filippini\orcidlink{0000-0001-8217-6832}}
\affiliation{Department of Physics, University of Illinois at Urbana-Champaign, Urbana, IL 61801, USA}
\affiliation{Department of Astronomy, University of Illinois at Urbana-Champaign, Urbana, IL 61801, USA}

\author{A.~Fortes}
\affiliation{Department of Physics, Stanford University, Stanford, CA 94305, USA}

\author{M.~Gao}
\affiliation{Department of Physics, California Institute of Technology, 1200 E. California Boulevard, Pasadena, CA 91125, USA}

\author{C.~Giannakopoulos}
\affiliation{Department of Physics, University of Cincinnati, Cincinnati, OH 45221, USA}

\author{N.~Goeckner-Wald}
\affiliation{Department of Physics, Stanford University, Stanford, CA 94305, USA}

\author{D.~C.~Goldfinger\orcidlink{0000-0001-5268-8423}}
\affiliation{Department of Physics, Stanford University, Stanford, CA 94305, USA}

\author{S.~Gratton}
\affiliation{Centre for Theoretical Cosmology, DAMTP, University of Cambridge, Wilberforce Road, Cambridge CB3 0WA, UK}
\affiliation{Kavli Institute for Cosmology Cambridge, Madingley Road, Cambridge CB3 0HA, UK}

\author{J.~A.~Grayson}
\affiliation{Department of Physics, Stanford University, Stanford, CA 94305, USA}

\author{A.~Greathouse}
\affiliation{Department of Physics, California Institute of Technology, 1200 E. California Boulevard, Pasadena, CA 91125, USA}

\author{P.~K.~Grimes\orcidlink{0000-0001-9292-6297}}
\affiliation{Center for Astrophysics, Harvard \& Smithsonian, Cambridge, MA 02138, USA}

\author{G.~Hall}
\affiliation{Minnesota Institute for Astrophysics, University of Minnesota, Minneapolis, MN 55455, USA}
\affiliation{Department of Physics, Stanford University, Stanford, CA 94305, USA}

\author{G.~Halal\orcidlink{0000-0003-2221-3018}}
\affiliation{Department of Physics, Stanford University, Stanford, CA 94305, USA}

\author{M.~Halpern}
\affiliation{Department of Physics and Astronomy, University of British Columbia, Vancouver, British Columbia, V6T 1Z1, Canada}

\author{E.~Hand}
\affiliation{Department of Physics, University of Cincinnati, Cincinnati, OH 45221, USA}

\author{S.~A.~Harrison}
\affiliation{Center for Astrophysics, Harvard \& Smithsonian, Cambridge, MA 02138, USA}

\author{S.~Henderson}
\affiliation{Kavli Institute for Particle Astrophysics and Cosmology, Stanford University, Stanford, CA 94305, USA}
\affiliation{SLAC National Accelerator Laboratory, Menlo Park, CA 94025, USA}

\author{T.~D.~Hoang}
\affiliation{School of Physics and Astronomy, University of Minnesota, Minneapolis, MN 55455, USA}

\author{J.~Hubmayr}
\affiliation{National Institute of Standards and Technology, Boulder, CO 80305, USA}

\author{H.~Hui\orcidlink{0000-0001-5812-1903}}
\affiliation{Department of Physics, California Institute of Technology, 1200 E. California Boulevard, Pasadena, CA 91125, USA}

\author{K.~D.~Irwin}
\affiliation{Department of Physics, Stanford University, Stanford, CA 94305, USA}

\author{J.~H.~Kang\orcidlink{0000-0002-3470-2954}}
\affiliation{Department of Physics, California Institute of Technology, 1200 E. California Boulevard, Pasadena, CA 91125, USA}

\author{K.~S.~Karkare\orcidlink{0000-0002-5215-6993}}
\affiliation{Department of Physics, Boston University, Boston, MA 02215, USA}
\affiliation{Kavli Institute for Particle Astrophysics and Cosmology, Stanford University, Stanford, CA 94305, USA}
\affiliation{SLAC National Accelerator Laboratory, Menlo Park, CA 94025, USA}

\author{S.~Kefeli}
\affiliation{Department of Physics, California Institute of Technology, 1200 E. California Boulevard, Pasadena, CA 91125, USA}

\author{J.~M.~Kovac\orcidlink{0009-0003-5432-7180}}
\affiliation{Department of Physics, Harvard University, Cambridge, MA 02138, USA}
\affiliation{Center for Astrophysics, Harvard \& Smithsonian, Cambridge, MA 02138, USA}

\author{C.~Kuo}
\affiliation{Department of Physics, Stanford University, Stanford, CA 94305, USA}

\author{K.~Lasko\orcidlink{0000-0002-4540-1495}}
\affiliation{School of Physics and Astronomy, University of Minnesota, Minneapolis, MN 55455, USA}
\affiliation{Minnesota Institute for Astrophysics, University of Minnesota, Minneapolis, MN 55455, USA}

\author{K.~Lau\orcidlink{0000-0002-6445-2407}}
\affiliation{Department of Physics, California Institute of Technology, 1200 E. California Boulevard, Pasadena, CA 91125, USA}

\author{M.~Lautzenhiser}
\affiliation{Department of Physics, University of Cincinnati, Cincinnati, OH 45221, USA}

\author{A.~Lennox}
\affiliation{Department of Astronomy, University of Illinois at Urbana-Champaign, Urbana, IL 61801, USA}

\author{T.~Liu\orcidlink{0000-0001-5677-5188}}
\affiliation{Department of Physics, Stanford University, Stanford, CA 94305, USA}

\author{S.~Mackey}
\affiliation{Kavli Institute for Cosmological Physics, University of Chicago, Chicago, IL 60637, USA}

\author{N.~Maher}
\affiliation{School of Physics and Astronomy, University of Minnesota, Minneapolis, MN 55455, USA}

\author{K.~G.~Megerian}
\affiliation{Jet Propulsion Laboratory, California Institute of Technology, Pasadena, CA 91109, USA}

\author{L.~Minutolo}
\affiliation{Department of Physics, California Institute of Technology, 1200 E. California Boulevard, Pasadena, CA 91125, USA}

\author{L.~Moncelsi\orcidlink{0000-0002-4242-3015}}
\affiliation{Department of Physics, California Institute of Technology, 1200 E. California Boulevard, Pasadena, CA 91125, USA}

\author{Y.~Nakato}
\affiliation{Department of Physics, Stanford University, Stanford, CA 94305, USA}

\author{H.~T.~Nguyen}
\affiliation{Jet Propulsion Laboratory, California Institute of Technology, Pasadena, CA 91109, USA}
\affiliation{Department of Physics, California Institute of Technology, 1200 E. California Boulevard, Pasadena, CA 91125, USA}

\author{R.~O'Brient}
\affiliation{Jet Propulsion Laboratory, California Institute of Technology, Pasadena, CA 91109, USA}
\affiliation{Department of Physics, California Institute of Technology, 1200 E. California Boulevard, Pasadena, CA 91125, USA}

\author{S.~N.~Paine}
\affiliation{Center for Astrophysics, Harvard \& Smithsonian, Cambridge, MA 02138, USA}

\author{A.~Patel}
\affiliation{Department of Physics, California Institute of Technology, 1200 E. California Boulevard, Pasadena, CA 91125, USA}

\author{M.~A.~Petroff\orcidlink{0000-0002-4436-4215}}
\affiliation{Center for Astrophysics, Harvard \& Smithsonian, Cambridge, MA 02138, USA}

\author{A.~R.~Polish\orcidlink{0000-0002-7822-6179}}
\affiliation{Department of Physics, Harvard University, Cambridge, MA 02138, USA}
\affiliation{Center for Astrophysics, Harvard \& Smithsonian, Cambridge, MA 02138, USA}

\author{T.~Prouve}
\affiliation{Service des Basses Temperatures, Commissariat a l’Energie Atomique, 38054 Grenoble, France}

\author{C.~Pryke\orcidlink{0000-0003-3983-6668}}
\affiliation{School of Physics and Astronomy, University of Minnesota, Minneapolis, MN 55455, USA}

\author{C.~D.~Reintsema}
\affiliation{National Institute of Standards and Technology, Boulder, CO 80305, USA}

\author{S.~Richter}
\affiliation{Center for Astrophysics, Harvard \& Smithsonian, Cambridge, MA 02138, USA}

\author{T.~Romand}
\affiliation{Department of Physics, California Institute of Technology, 1200 E. California Boulevard, Pasadena, CA 91125, USA}

\author{M.~Salatino}
\affiliation{Department of Physics, Stanford University, Stanford, CA 94305, USA}

\author{A.~Schillaci}
\affiliation{Department of Physics, California Institute of Technology, 1200 E. California Boulevard, Pasadena, CA 91125, USA}

\author{B.~Schmitt}
\affiliation{Center for Astrophysics, Harvard \& Smithsonian, Cambridge, MA 02138, USA}

\author{R.~Schwarz}
\affiliation{School of Physics and Astronomy, University of Minnesota, Minneapolis, MN 55455, USA}

\author{C.~D.~Sheehy}
\affiliation{School of Physics and Astronomy, University of Minnesota, Minneapolis, MN 55455, USA}

\author{B.~Singari\orcidlink{0000-0001-7387-0881}}
\affiliation{School of Physics and Astronomy, University of Minnesota, Minneapolis, MN 55455, USA}
\affiliation{Minnesota Institute for Astrophysics, University of Minnesota, Minneapolis, MN 55455, USA}%

\author{A.~Soliman}
\affiliation{Jet Propulsion Laboratory, California Institute of Technology, Pasadena, CA 91109, USA}
\affiliation{Department of Physics, California Institute of Technology, 1200 E. California Boulevard, Pasadena, CA 91125, USA}

\author{T.~St Germaine}
\affiliation{Center for Astrophysics, Harvard \& Smithsonian, Cambridge, MA 02138, USA}

\author{A.~Steiger\orcidlink{0000-0003-0260-605X}}
\affiliation{Department of Physics, California Institute of Technology, 1200 E. California Boulevard, Pasadena, CA 91125, USA}

\author{B.~Steinbach}
\affiliation{Department of Physics, California Institute of Technology, 1200 E. California Boulevard, Pasadena, CA 91125, USA}

\author{R.~Sudiwala}
\affiliation{School of Physics and Astronomy, Cardiff University, Cardiff, CF24 3AA, United Kingdom}

\author{G.~P.~Teply}
\affiliation{Department of Physics, California Institute of Technology, 1200 E. California Boulevard, Pasadena, CA 91125, USA}

\author{K.~L.~Thompson}
\affiliation{Department of Physics, Stanford University, Stanford, CA 94305, USA}

\author{C.~Tucker\orcidlink{0000-0002-1851-3918}}
\affiliation{School of Physics and Astronomy, Cardiff University, Cardiff, CF24 3AA, United Kingdom}

\author{A.~D.~Turner}
\affiliation{Jet Propulsion Laboratory, California Institute of Technology, Pasadena, CA 91109, USA}

\author{C.~Vergès\orcidlink{0000-0002-3942-1609}}
\affiliation{Physics Division, Lawrence Berkeley National Laboratory, Berkeley, CA 94720, USA}
\affiliation{Center for Astrophysics, Harvard \& Smithsonian, Cambridge, MA 02138, USA}

\author{A.~G.~Vieregg}
\affiliation{Kavli Institute for Cosmological Physics, University of Chicago, Chicago, IL 60637, USA}

\author{A.~Wandui\orcidlink{0000-0002-8232-7343}}
\affiliation{Department of Physics, California Institute of Technology, 1200 E. California Boulevard, Pasadena, CA 91125, USA}

\author{A.~C.~Weber}
\affiliation{Jet Propulsion Laboratory, California Institute of Technology, Pasadena, CA 91109, USA}

\author{J.~Willmert\orcidlink{0000-0002-6452-4693}}
\affiliation{School of Physics and Astronomy, University of Minnesota, Minneapolis, MN 55455, USA}

\author{C.~L.~Wong}
\affiliation{Center for Astrophysics, Harvard \& Smithsonian, Cambridge, MA 02138, USA}
\affiliation{Department of Physics, Harvard University, Cambridge, MA 02138, USA}

\author{W.~L.~K.~Wu\orcidlink{0000-0001-5411-6920}}
\affiliation{SLAC National Accelerator Laboratory, Menlo Park, CA 94025, USA}
\affiliation{Kavli Institute for Particle Astrophysics and Cosmology, Stanford University, Stanford, CA 94305, USA}
\affiliation{Department of Physics, Stanford University, Stanford, CA 94305, USA}

\author{H.~Yang}
\affiliation{Department of Physics, Stanford University, Stanford, CA 94305, USA}

\author{C.~Yu\orcidlink{0000-0002-8542-232X}}
\affiliation{Kavli Institute for Cosmological Physics, University of Chicago, Chicago, IL 60637, USA}

\author{L.~Zeng\orcidlink{0000-0001-6924-9072}}
\affiliation{Center for Astrophysics, Harvard \& Smithsonian, Cambridge, MA 02138, USA}

\author{C.~Zhang\orcidlink{0000-0001-8288-5823}}
\affiliation{Department of Physics, Stanford University, Stanford, CA 94305, USA}

\author{S.~Zhang}
\affiliation{Department of Physics, California Institute of Technology, 1200 E. California Boulevard, Pasadena, CA 91125, USA}

\date{\today}

\begin{abstract}
We present component-separated polarization maps of the cosmic microwave background (CMB) and Galactic thermal dust emission, derived using data from the BICEP/\textit{Keck} experiments through the 2018 observing season and \textit{Planck}. By employing a maximum-likelihood method that utilizes observing matrices, we produce unbiased maps of the CMB and dust signals. We outline the computational challenges and demonstrate an efficient implementation of the component map estimator. We show methods to compute and characterize power spectra of these maps, opening up an alternative way to infer the tensor-to-scalar ratio from our data. We compare the results of this map-based separation method with the baseline BICEP/\textit{Keck} analysis. Our analysis demonstrates consistency between the two methods, finding an 84\% correlation between the pipelines.
\end{abstract}

\maketitle

\section{Introduction}
Measurements of the CMB have been transformative for our understanding of the Universe, leading to the establishment of the $\Lambda$CDM mode as the standard cosmological paradigm. These successes were achieved through technological and analytical advances that overcame challenges posed by the faintness of the signal of interest: The detection of one part in $10^{-5}$ total intensity fluctuations on top of the isotropic thermal radiation at $T\approx2.7~\mathrm{K}$ \cite{Smoot1992} and the detection of CMB polarization, another two orders of magnitude smaller than the total intensity fluctuations \cite{Kovac2002}.\\

Significant effort has been made towards the next milestone in CMB measurements, the search for large-scale B-mode polarization at the nanokelvin level. This challenging measurement not only requires instruments with unprecedented sensitivity at millimeter wavelengths, but also experimental and analytical methods to remove astrophysical signals now approaching two orders of magnitude larger than the sought-after signal: Galactic foregrounds and weak gravitational lensing.\\

The most sensitive constraint on degree-scale B-mode polarization comes from the BICEP/\textit{Keck} collaboration \cite{BK18}, referred to as ``BK18" in the following, setting a 95\% upper-limit on the level of primordial B-mode power approximately one order of magnitude below the power of Galactic dust emission at 150 GHz within the observed patch. This is achieved by jointly fitting for an amplitude of primordial B-modes and seven foreground parameters following a model of the frequency and spatial behavior of Galactic dust and synchrotron emission. This likelihood, based on Ref.~\cite{Hamimeche2008}, which compares data and theory at the level of power spectra, or bandpowers, is widely used in current and upcoming analyses of experiments \cite{BKP,Planck2020spectra,Abazajian2022,Wolz2023}. \\

In this paper we explore an alternative approach to separating different sky components in real BICEP/\textit{Keck} data with the goal of producing maps of the polarized CMB and Galactic dust emission. We investigate the differences and consistency with the baseline approach of modeling foregrounds and the CMB at the level of power spectra. \\

\section{Map-Based Component Separation}
We first introduce the likelihood formalism underpinning our component-separation method. In a manner similar to the pixel-based spectral-fitting approaches of Refs.~\cite{Eriksen2006,PlanckCompsep,Stompor2009}, this framework produces unbiased maps of the CMB and dust signals. \\

\subsection{Observation model}
The component maps we produce are maximum likelihood estimates given the input of a data vector, $\mathbf{d}$. In our case, $\mathbf{d}$ is a stack of $N$ frequency-domain polarization maps ($Q$ and $U$ for each frequency)
\begin{equation}
    \mathbf{d}=
    \begin{pmatrix}
        \tilde{\mathbf{m}}_{1}\\
        \vdots \\
        \tilde{\mathbf{m}}_{{N}}\\
    \end{pmatrix}.
\end{equation}
In this notation $\tilde{\mathbf{m}}$ contains column vectors of $Q$ and $U$ maps stacked together
\begin{equation}
    \tilde{\mathbf{m}}=
    \begin{pmatrix}
        \tilde{\mathbf{m}}^{Q}\\
        \tilde{\mathbf{m}}^{U}
    \end{pmatrix},
\end{equation}
so $\tilde{\mathbf{m}}$ is a vector of length $2n_p$, where $n_p$ is the number of pixels, which is generally $\mathcal{O}(10^5)$.\\

The timestream filtering and deprojection of BICEP/\textit{Keck} frequency maps, $\tilde{\mathbf{m}}_{\nu}$, can be modeled as a linear operation represented by the observing matrix $\mathbf{R}$. This matrix is constructed alongside other BICEP/\textit{Keck} data products and plays an essential role in achieving sufficient B-mode purity to infer $r$ from B-mode power spectra \cite{bkmatrix}. This matrix can act on beam-convolved input maps to produce simulated maps as they would be observed by a BICEP/\textit{Keck} frequency band $\nu$
\begin{equation}
    \tilde{\mathbf{m}}_\nu=
    \mathbf{R}\mathbf{B}
    \mathbf{m}_\nu+ \mathbf{n}_\nu,
\end{equation}
where $\mathbf{B}$ is an operator convolving the $Q$ and $U$ maps with their respective symmetric beam functions, $b_\ell$. Both observing matrix and beam function can be different from frequency channel to frequency channel. Further, each simulation of a BICEP/\textit{Keck} $Q$ and $U$ frequency map includes an additive noise component, $\mathbf{n}$.

\subsection{Multifrequency model}

For previous BICEP/\textit{Keck} results, the baseline multi-frequency model assumes three components: CMB, Galactic dust, and Galactic synchrotron emission \cite{BKP,biceptwoX}. Given that the current data set is limited to WMAP and \textit{Planck} observations at low frequencies and measures an amplitude of polarized synchrotron emission consistent with zero in the observation patch, we only attempt to explicitly separate CMB and Galactic thermal dust in the present paper. We employ a parametric model for dust emission as a function of frequency, such that we can model each frequency map as a linear combination of CMB and dust
\begin{align}
    \mathbf{m}=
    \begin{pmatrix}
        \mathbf{m}_{1}\\
        \vdots \\
        \mathbf{m}_{N}\\
    \end{pmatrix}
    &=
    \begin{pmatrix}
        \mathbf{1} & \mathbf{f}_{1}\\
        \vdots & \vdots \\
        \mathbf{1} & \mathbf{f}_{N}\\
    \end{pmatrix}
    \begin{pmatrix}
        \mathbf{s}^\textrm{CMB} \\
        \mathbf{s}^\textrm{dust} 
    \end{pmatrix}
    \equiv\mathbf{F} \mathbf{s},
\end{align}
where $\mathbf{1}$ is the $2n_p\times2n_p$ identity matrix and $\mathbf{f}_i\equiv f_i\mathbf{1}$ are scaling factors multiplied to that identity matrix. We assume a modified black-body spectrum for Galactic dust such that the scaling factor for the $i$th frequency band is given by
\begin{equation}
    f_i \sim \int b_i(\nu) \frac{\nu^{3+\beta_d}}{\exp\left(\frac{h\nu}{k_B T_d}\right)-1} d\nu,
\end{equation}
where $b_i(\nu)$ is the respective bandpass. In this work, we assume a spatially constant $\beta_d$ and do not treat it as a free parameter in our maximum-likelihood search.\\

The end result is a model relating sky component maps to observed BICEP/\textit{Keck} and external frequency maps is given by
\begin{equation}
    \tilde{\mathbf{m}}=\mathbf{R}\mathbf{B}\mathbf{F} \mathbf{s}+\mathbf{n} \equiv \mathbf{A} \mathbf{s}+\mathbf{n}.
\end{equation}

\subsection{Noise model}

Apart from the signal, each frequency map also includes an additive noise term. One simple and straightforward choice is to characterize the noise fluctuations after timestream filtering and binning into maps as Gaussian random fluctuations, which can therefore be fully described in terms of their noise covariance matrix
\begin{equation}
    \hat{\mathbf{N}} = \left\langle \mathbf{n} \mathbf{n}^T \right\rangle.
\end{equation}
In this work we approximate this matrix to be diagonal in pixel space, with the variance maps of each frequency channel populating its diagonal. The off-diagonal noise correlations in map-space are on the 5\% level and are neglected here. Note that a mismatch between the true noise covariance and this estimate will not cause a biased estimate as long as the noise is Gaussian. However, it can make the estimator sub-optimal.\\

\subsection{Likelihood formalism}
We start out with the problem of maximizing the \textit{full likelihood} 
\begin{equation}
    -2\log P(\mathbf{s},\beta_d|\mathbf{d}) = (\mathbf{d}-\mathbf{A}(\beta_d)\mathbf{s})^T \hat{\mathbf{N}}^{-1} (\mathbf{d}-\mathbf{A}(\beta_d)\mathbf{s}),
    \label{eq:flatposterior}
\end{equation}
where both the foreground model, represented by the parameter $\beta_d$, and the sky signal $\mathbf{s}$ are unknown. We will make use of the maximization of the profile likelihood $-2\log P(\mathbf{s},\beta_d=\beta_d^*|\mathbf{d})$, where we fix the parameter in the foreground model in $\mathbf{A}$ to the value $\beta_d^*$. The maximization of this profile likelihood yields the generalized least squares estimator \cite{Aitken_1936}
\begin{equation}
    \hat{\mathbf{s}} = \left( \mathbf{A}^T \hat{\mathbf{N}}^{-1} \mathbf{A} \right)^{-1} \mathbf{A}^T \hat{\mathbf{N}}^{-1} \mathbf{d}.
    \label{eq:mapskysignal}
\end{equation}
This makes the problem of separating out components a computationally expensive problem, since the inversion of the large system matrix $\mathbf{A}^T \hat{\mathbf{N}}^{-1} \mathbf{A}$ is expensive.\\

\subsection{Foreground model}

\begin{figure}
    \centering
    \includegraphics[width=\linewidth]{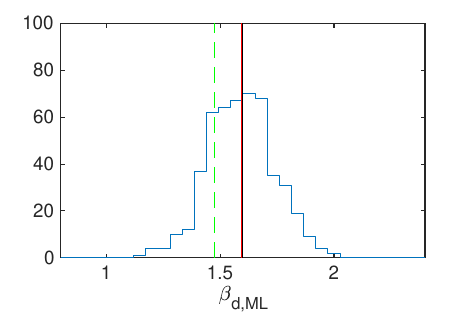}
    \caption{Histogram of the maximum likelihood values of the dust spectral index $\beta_d$ obtained in the baseline BK18 \cite{BK18} auto-/cross-spectrum analysis. The black vertical line indicates the input value to the simulations, the (nearly coincident) red line indicates the mean of the recovered best-fit values, and the dashed green line marks the best-fit value of the real data.}
    \label{fig:betad}
\end{figure}

This study leverages the best-fit foreground parameters derived from the mainline BK18 analysis \cite{BK18}. There we fit a model of CMB, dust, and synchrotron to auto- and cross-frequency spectra of the real data and 499 simulated skies, including CMB, dust, and instrumental noise. The distribution of the resulting best-fit values of the dust spectral index is shown in Fig.~\ref{fig:betad} and can be used to propagate the statistical uncertainty of this parameter to the component-separated maps. We use the respective best-fit value for $\beta_d$ for each simulation realization in the component map estimator in Eq. \ref{eq:mapskysignal}.\\

Possible extensions of this approach would be the inclusion of a consistent map-based fitting of $\beta_d$ and $\beta_s$ in a dedicated two-step approach following for example Ref.~\cite{Stompor2009, de_Belsunce_2022}. This formalism also more directly allows for fitting spatially varying foreground parameters at map level, which would make this pipeline more robust against biases due to complex foregrounds. Moreover, Refs.~\cite{Leloup2023,Morshed2024} have recently introduced and implemented a non-parametric maximum-likelihood framework for CMB foreground separation. We leave an exploration of these options for future work.\\

\subsection{Comparison to the BICEP/\textit{Keck} baseline foreground cleaning}\label{sec:pipecomparison}

The baseline $r$-analysis of BICEP/\textit{Keck} accounts for foreground contamination at the power-spectrum-level by modeling the foreground bias from dust and synchrotron using a spatial and frequency-space model. In the likelihood framework above this corresponds to adding a signal prior to the full likelihood in Eq.~\ref{eq:flatposterior}
\begin{equation}
    -2\log P(\mathbf{s},\beta_d|\mathbf{d}) = ... + \mathbf{s}^T \mathbf{S}^{-1} \mathbf{s}  + \log\det \mathbf{S}.
    \label{eq:signalprior}
\end{equation}

Here, $\mathbf{S}$ represents the signal covariance matrix, which incorporates the auto-correlations of CMB, dust, and synchrotron. These components are modeled in harmonic space using parametric representations of their respective power spectra. Marginalizing this likelihood over the sky signal $\mathbf{s}$ and applying the arguments of Ref.~\cite{Hamimeche2008} yields the baseline BK18 auto- and cross-spectra likelihood. \\

In practice, modeling the data at the level of multi-frequency spectra can lead to complications. In BK18, for example, the analysis has to account for two significantly different observation footprints caused by the difference in field-of-view between BICEP2/\textit{Keck} and BICEP3. This was solved by computing cross-spectra between the maps using a small-field mask for BICEP2/\textit{Keck} maps and a large-field mask for BICEP3 and external maps \cite{BK18}. \\

Furthermore, introducing a prior on the foreground behavior may result in biased foreground parameter estimates. In practice, the traditional prior on the foreground behavior in harmonic space is a power-law power spectrum with an amplitude and spectral slope. While measurements of thermal dust emission and synchrotron emission suggest that this assumption is approximately true, failure to capture the actual shape of the bandpowers dust and synchrotron happen to produce in our patch can lead to misleading constraints on foreground spectral parameters. Hence, an alternative pipeline can improve robustness and build confidence in our foreground-cleaning capability.\\

\section{Data}\label{sec:dataandsims}
The BK18 dataset consists of observations from the BICEP2, \textit{Keck Array}, and BICEP3 receivers located at the South Pole Station in Antarctica. The \biceptwo\ receiver observed at 150\,GHz from 2010--2012~\cite{biceptwoII} with $\approx 500$ bolometric detectors. The \keckarray\ consisting of five copies of \biceptwo{-size} receivers running from 2012--2019, initially observed at 150\,GHz but switched over time to 95 and 220\,GHz~\cite{biceptwoV}. \bicepthree\ is a single, scaled up receiver at $95$ GHz which started science observations in 2016~\cite{bicepthree} with $\approx 2500$ detectors.\\

\biceptwo\ and \keckarray\ both mapped a region of sky centered at RA 0h, Dec.\ $-57.5\deg$ with an effective area of $\approx 400$ square degrees \cite{biceptwoX}. \bicepthree\ has a larger instantaneous field of view and hence naturally maps a larger sky area with an effective area of $\approx 600$ square degrees. This results in small-field maps at 95/150/220\,GHz and a large-field map at 95\,GHz. In this paper we make use of the standard BK18 maps and simulations.\\

These maps were produced with a filter-and-bin map-maker, removing a third-order polynomial, scan-synchronous-signal, and T-to-P-leakage templates from the data before accumulating it into maps of equirectangular pixelization, with $0.25\deg$ square pixels at declination $-57.5\deg$. The map binning operation weights by the inverse variance of the timestream data. These weights, binned into maps, are used to get an estimate of the noise variance in $T$, $Q$, and $U$.\\

Additionally we include external data to leverage the high-frequency observations of the \textit{Planck} satellite mission covering the entire sky. In this paper we make use of the 100, 143, 217, and 353 GHz $Q$ and $U$ maps from the NPIPE processing of the \textit{Planck} data \cite{PlanckNPIPE}. As opposed to the mainline analysis of BK18, these maps are not reobserved prior to using them in the estimator.\\

\section{Implementation}

\subsection{Maximum-likelihood estimator}
\begin{figure*}
    \centering
    \includegraphics[width=\linewidth]{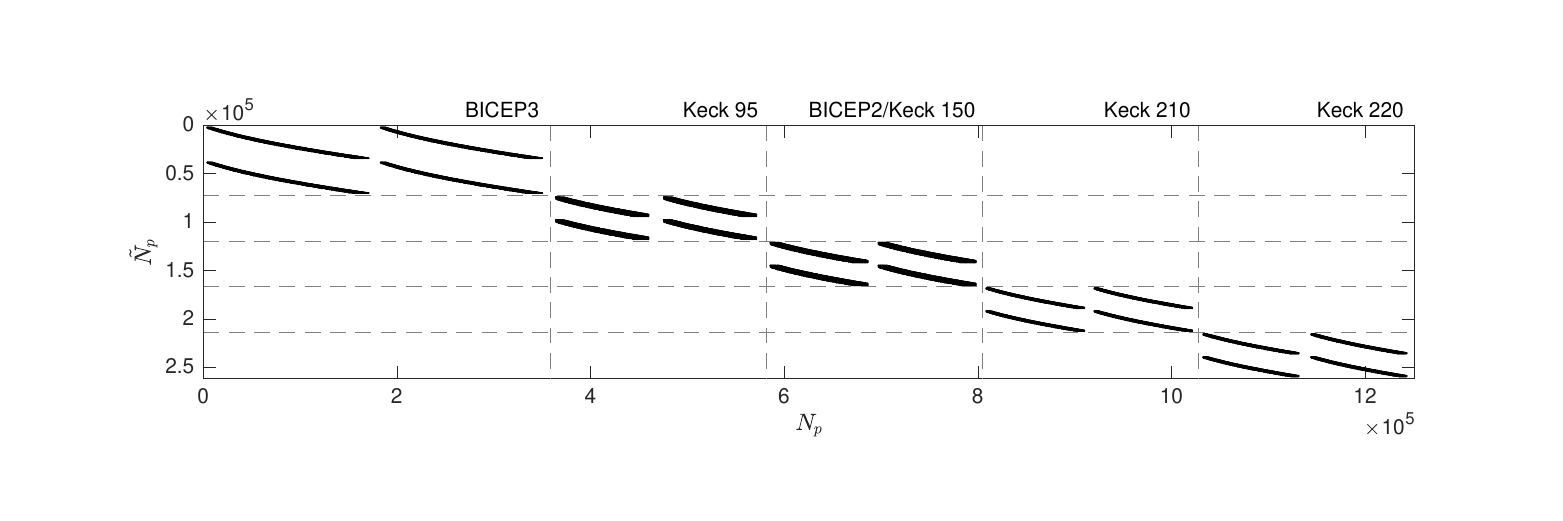}
    \caption{Plot of the non-zero matrix elements of the \bicep/\keck\ internal maps' observing matrices, i.e., the \bicep/\keck-specific block of the matrix $\mathbf{R}$ used in this work. Applying a vector including beam-convolved $Q$ \& $U$ maps to the right of this matrix results in a vector of filtered $Q$ \& $U$ maps in the flat pixelization used in BK18. The top-left block corresponds to the BICEP3 map, while the following blocks along the diagonal contain the observing matrices for the \keck\ 95\,GHz, \biceptwo/\keck\ 150\,GHz, \keck\ 210\,GHz, and \keck\ 220\,GHz channels. Within each block on the diagonal, the top row produces a \bicep/\keck\ $Q$ map and the bottom row a \bicep/\keck\ $U$ map from a vector of stacked $Q$ \& $U$ maps. This part of the matrix contains about 5 billion non-zero elements and is the biggest computational challenge in this analysis.}
    \label{fig:rmatrix}
\end{figure*}

The primary computational challenge is solving the large linear system in the estimator in Eq.~\ref{eq:mapskysignal} via matrix inversion. We employ an iterative method to numerically solve for $\hat{\mathbf{s}}$ in the linear equation
\begin{equation}
    \left( \mathbf{F}^T\mathbf{B}^T\mathbf{R}^T \hat{\mathbf{N}}^{-1} \mathbf{R}\mathbf{B}\mathbf{F} \right) \hat{\mathbf{s}} = \mathbf{F}^T\mathbf{B}^T\mathbf{R}^T \hat{\mathbf{N}}^{-1} \mathbf{d}.
    \label{eq:mapskysignalpcg}
\end{equation}
This method allows us to never have to explicitly construct the large $4n_p \times 4n_p$ matrix on the left-hand side. It is merely sufficient to build a routine that applies the matrix to the solution vector $\hat{\mathbf{s}}$. This leads to a significant speed-up given that $\mathbf{R}$, $\mathbf{B}$ and $\mathbf{F}$ are sparse operators either in pixel or harmonic spaces and thanks to fast routines available to transform between the two spaces. Fig.~\ref{fig:rmatrix} highlights the sparsity pattern of the internal BICEP/\textit{Keck} observing matrices, showing the blocks corresponding to the BICEP/\textit{Keck} frequency bands. This matrix is heavily concentrated at the sub-diagonals within each frequency-channel $Q$ and $U$ blocks. The beam convolution operator $\mathbf{B}$ convolves a map with a beam function, which is a simple multiplication in harmonic space and should hence be performed after spherically-transforming the map. In this work, we assume no spatial variation of the foregrounds and hence the $\mathbf{F}$-matrix is simple in either space. Such variations could be easily incorporated, however, as long as they are diagonal in pixel or harmonic space.\\

We solve Eq.~\ref{eq:mapskysignal} with a preconditioned iterative method \cite{pcg}, employing a block Jacobi preconditioner. The use of a preconditioner significantly speeds up the convergence of the iterative method by approximating the linear operator with an easily invertible matrix. We construct this preconditioner by ignoring the observing matrix and beam in the system matrix
\begin{equation}
    \mathbf{M}\equiv\mathbf{F}^T \hat{\mathbf{N}}^{-1} \mathbf{F}.
\end{equation}
Adding external data from \textit{Planck} not only extends the frequency coverage but also stabilizes the numerical problem by filling in modes that are in the null-space of the BICEP/\textit{Keck} observing matrices.

\subsection{Numerical experiments}

To focus on the problem of inverting the system matrix on the right-hand side of Eq.~\ref{eq:mapskysignal}, we perform numerical experiments on a more simplified problem: we correct for the effects of filtering and deprojection at the map level and produce ``unbiased" maps for a single frequency only. This effectively amounts to inverting a single-frequency observing matrix by solving the estimator, e.g., for BICEP3\,($B3$) only
\begin{equation}
    \hat{\mathbf{s}}_{B3} =  \left( \mathbf{R}^T \hat{\mathbf{N}}_{B3}^{-1} \mathbf{R} \right)^{-1}\mathbf{R}^T \hat{\mathbf{N}}_{B3}^{-1} \mathbf{d}_{B3}.
    \label{eq:B3mapskysignal}
\end{equation}

We know that this problem is mathematically ill-defined since the matrix $\mathbf{R}^T \hat{\mathbf{N}}^{-1} \mathbf{R}$ is strictly not invertible. We can regularize this problem by adding external data with more complete mode coverage than BICEP/\textit{Keck}, such as \textit{Planck}. For example, we can produce a combined map at around 100 GHz with BICEP3 and \textit{Planck} 100 GHz ($P$) data, assuming the latter has negligible filtering suppression, by solving the estimator 
\begin{equation}
    \hat{\mathbf{s}}_\textrm{95\ GHz} =  \left( \mathbf{R}^T \hat{\mathbf{N}}_{B3}^{-1} \mathbf{R} + \hat{\mathbf{N}}_{P}^{-1} \right)^{-1}\left(\mathbf{R}^T \hat{\mathbf{N}}_{B3}^{-1} \mathbf{d}_{B3} + \hat{\mathbf{N}}_{P}^{-1} \mathbf{d}_{P}\right).
    \label{eq:B3Pmapskysignal}
\end{equation}

\begin{figure*}
    \centering
    \includegraphics[width=\linewidth]{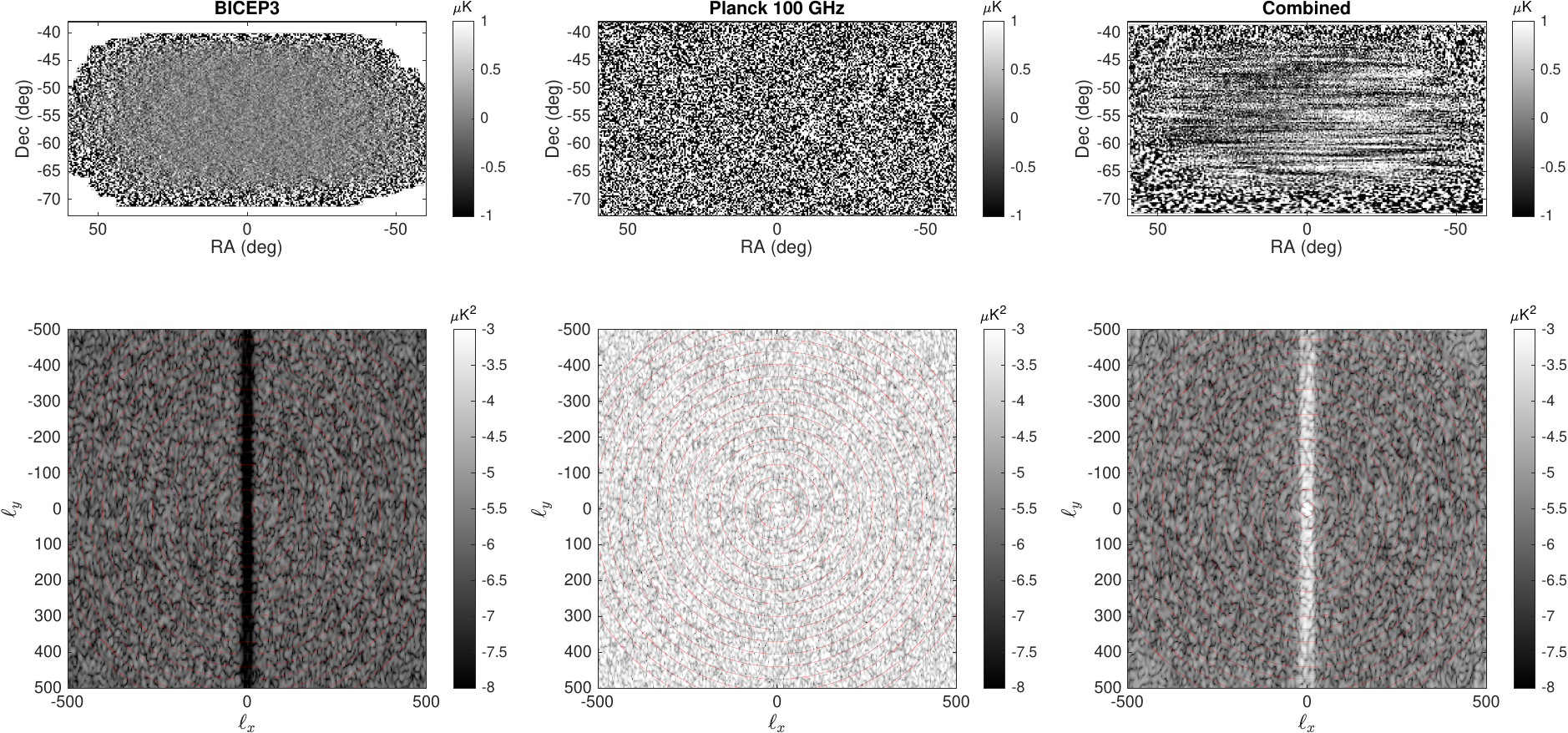}
    \caption{$Q$ maps (top row) and the logarithmic (base 10) $EE$ two-dimensional auto-power spectrum (bottom row) of one noise simulation for BICEP3 (left column), \textit{Planck} HFI 100 GHz (middle column), and their combination (right column). The combination estimator corrects for filtering suppression at the map-level and hence boosts the noise compared to the filtered BICEP3 map and fills in modes from \textit{Planck} for small $\ell_x$. This is why the combined $Q$ noise maps show strong horizontal stripes in the central BICEP3 map region.}
    \label{fig:combination}
\end{figure*}

In Fig.~\ref{fig:combination} we show the $Q$ maps and the $EE$ two-dimensional auto-power spectrum of a noise simulation for the BICEP3 and \textit{Planck} 100 GHz input maps, as well as their combination $\hat{\mathbf{s}}_\textrm{95\ GHz}$. It illustrates how this estimator fills in the filtered-out modes in the BICEP3 map with noisier \textit{Planck} modes in the ``poly-trench'' along the $\ell_y$ direction. This causes characteristic stripes in the combined map. \\

\begin{figure}
    \centering
    \includegraphics[width=\linewidth]{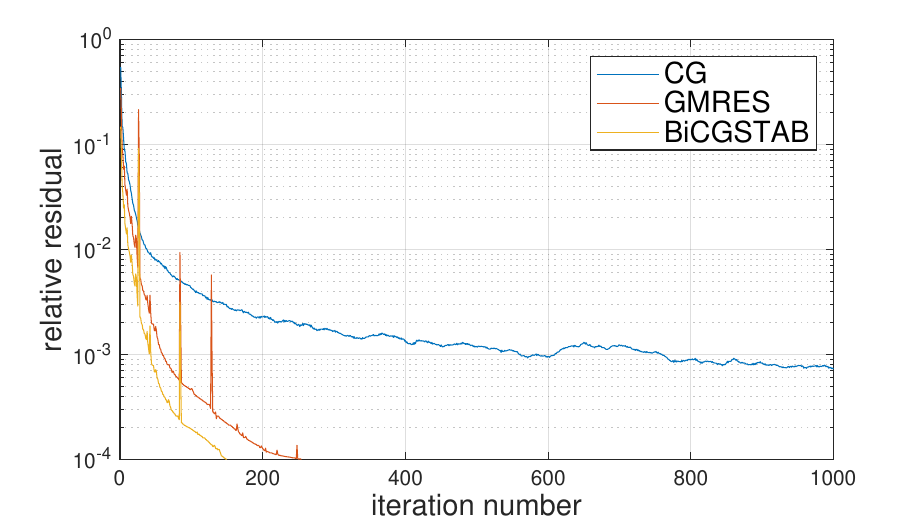}
    \caption{The relative residual $||\mathbf{A}\mathbf{s}-\mathbf{b}||/||\mathbf{b}||$, where $\mathbf{A}\equiv \mathbf{R}^T \hat{\mathbf{N}}_{B3}^{-1} \mathbf{R} + \hat{\mathbf{N}}_{P}^{-1}$ and $\mathbf{b}=\mathbf{R}^T \hat{\mathbf{N}}_{B3}^{-1} \mathbf{d}_{B3} + \hat{\mathbf{N}}_{P}^{-1} \mathbf{d}_{P}$, for each iteration of the preconditioned iterative method. We test three different iterative solvers: the classic CG, GMRES, and Bi-CGSTAB \cite{pcg}. The spikes are due to numerical noise, which these iterative solvers are susceptible to \cite{Sidje2011}.}
    \label{fig:pcgresidualmethods}
\end{figure}

In Fig.~\ref{fig:pcgresidualmethods} we explore three different preconditioned iterative solvers for this problem applied to signal-and-noise maps: the Conjugate Gradient method (CG), the Generalized Minimal Residual method (GMRES), and the Biconjugate Gradient Stabilized method (Bi-CGSTAB) \cite{pcg}. We find the latter to have superior convergence behavior and will use it in the following work. It should be noted that Bi-CGSTAB requires two matrix-vector multiplications per iteration, whereas GMRES requires only one. However, we find that the effective wall-clock time difference between the two is small, as the faster convergence rate of BiCGSTAB often compensates for the additional computational cost per step.\\

\begin{figure}
    \centering
    \includegraphics[width=\linewidth]{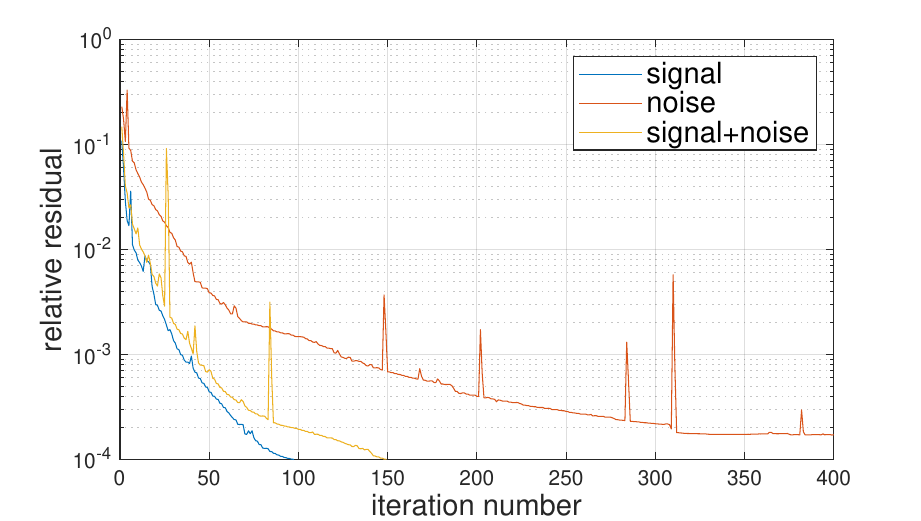}
    \caption{The relative residual as defined in Fig.~\ref{fig:pcgresidualmethods} for each iteration of the (Bi-CGSTAB) preconditioned conjugate gradient method. We show the convergence performance for a signal-only, noise-only, and a signal-and-noise simulation.}
    \label{fig:pcgresidual}
\end{figure}

In Fig.~\ref{fig:pcgresidual} we show the convergence behavior of the iterative solver for a signal-only, a noise-only, and a signal-and-noise simulation. Due to noise inhomogeneity, the iterative method struggles to converge. For a homogeneous input map such as a signal-only simulations, the iterative solver converges much faster.\\

\begin{figure}
    \centering
    \includegraphics[width=\linewidth]{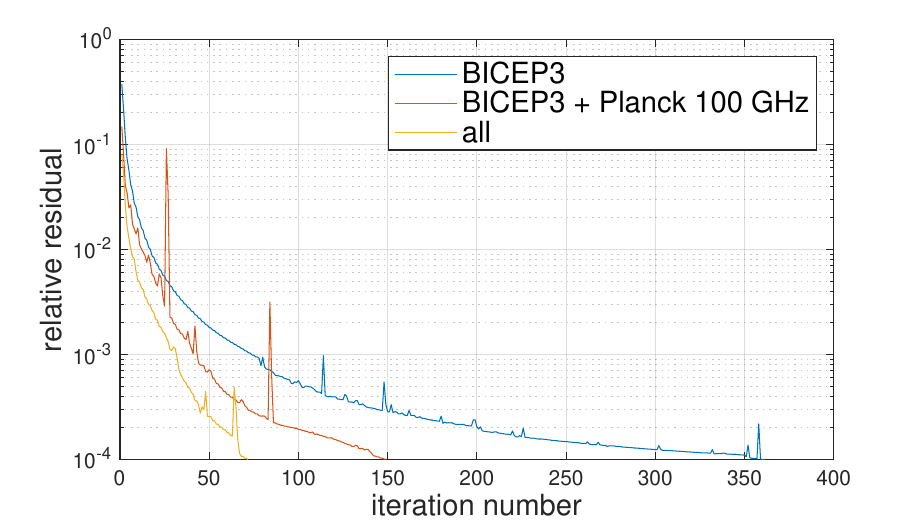}
    \caption{The relative residual as defined in the caption of Fig.~\ref{fig:pcgresidualmethods} for each PCG iteration step. We show a comparison to solve for the frequency-map solution for BICEP3 and the BICEP3+\textit{Planck} combination, as well as the CMB component map solution given all input frequency maps as described in Sec.~\ref{sec:dataandsims}.}
    \label{fig:pcgresidualcombinations}
\end{figure}

In Fig.~\ref{fig:pcgresidualcombinations} we show numerical experiments for a signal-and-noise simulation for different sizes of the input map vector. A single-frequency BICEP3 solution as in Eq.~\ref{eq:B3mapskysignal} will take a long time to converge, and a faithful estimate is not guaranteed due to the non-invertibility of the system matrix. The regularization with an external \textit{Planck} map as described in Eq.~\ref{eq:B3Pmapskysignal} decreases the number of iterations needed to reach the convergence criterion of a relative residual of $10^{-4}$. Lastly, the full problem of combining all frequency maps described in Sec.~\ref{sec:dataandsims} to solve for a CMB component map reaches this criterion in $\mathcal{O}(70)$ steps, noting that a larger data vector leads to larger runtime per iteration and higher memory requirement. 

\section{Noise Properties}

The output of the estimator, $\hat{\mathbf{s}}$, is an unbiased map of CMB and dust, meaning the signal suppression due to filtering and deprojection is corrected for at the map-level. This comes at the expense of elevated, inhomogeneous noise in the final map. \\

\begin{figure*}
    \centering
    \includegraphics[width=.49\linewidth]{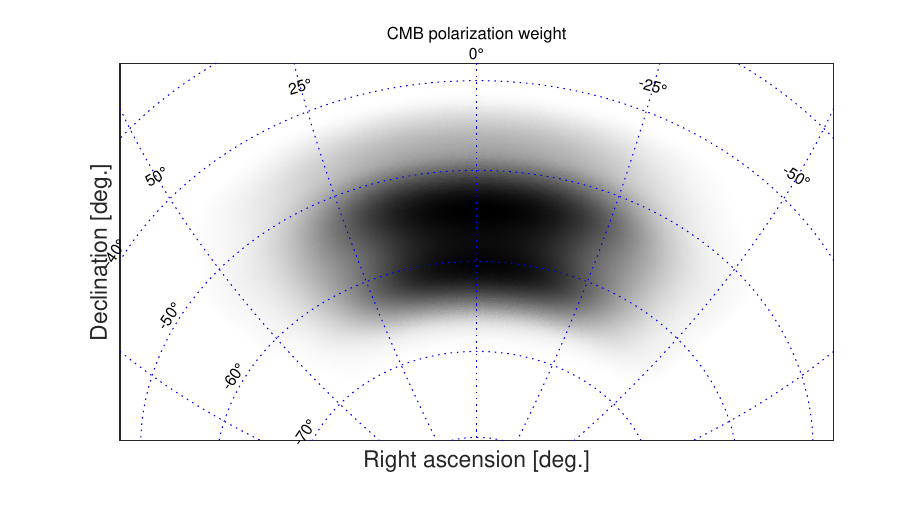}
    \includegraphics[width=.49\linewidth]{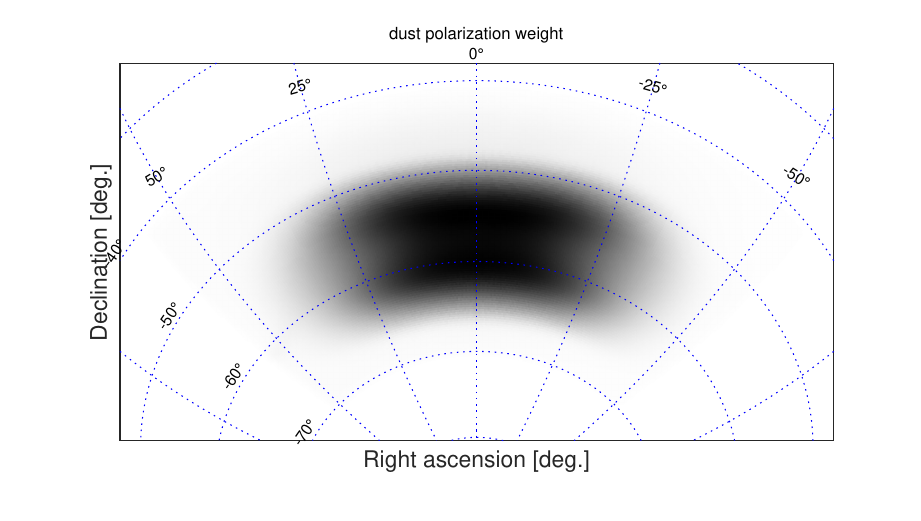}
    \caption{The polarization weight defined as the inverse arithmetic mean of the $Q$ and $U$ noise variance assuming white noise per pixel for the CMB component (left) and the dust component (right) in arbitrary units. Due to the extended coverage of BICEP3, the CMB extends above $50^\circ$ in declination, while the dust component map is most sensitive in the BICEP2/\textit{Keck} region.}
    \label{fig:mapvar}
\end{figure*}

We build a map-level weight from the $Q$ and $U$ variance maps produced in the BICEP/\textit{Keck} map-making. Using our fiducial foreground model, we can build approximate maps of the Q and U variance in the CMB and dust component maps by taking the diagonal of
\begin{equation}
\tilde{\mathbf{N}}=\left(\mathbf{F}^T\mathbf{N}^{-1}\mathbf{F}\right)^{-1}.
\end {equation}
We then weight the CMB and dust component maps by the inverse of the arithmetic mean of their $Q$ and $U$ variance maps, just like we do for the BICEP/\textit{Keck} frequency maps in Ref.~\cite{BK18}. These weights are shown in Fig.~\ref{fig:mapvar}. We observe that the CMB component weight extends to higher declinations due to the larger extent of the BICEP3 map, which significantly contributes to the CMB reconstruction. The dust map, however, is mostly limited to the \textit{Keck} 220 GHz patch.\\

\begin{figure}
    \centering
    \includegraphics[width=\linewidth]{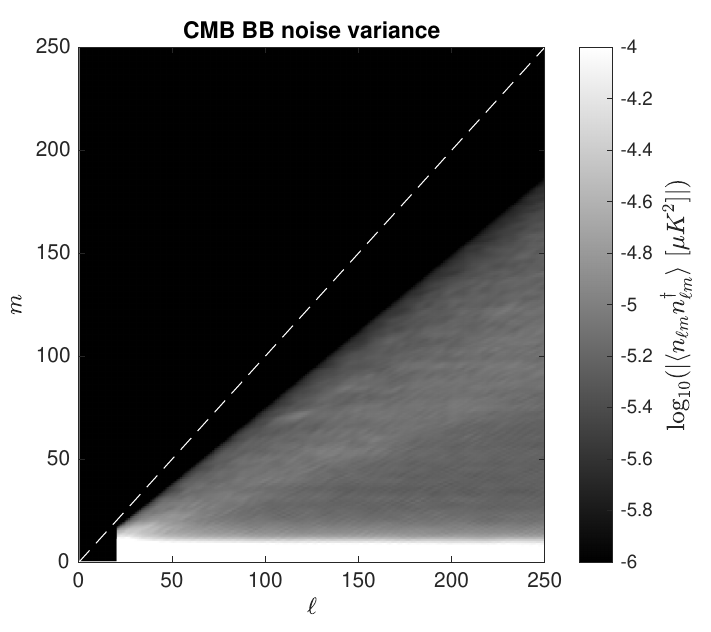}
    \caption{The noise variance per $(\ell,m)$ mode of the CMB component map obtained by averaging over 499 noise-only simulations, after applying the corresponding map-level weight shown in Fig.~\ref{fig:mapvar} before computing harmonic coefficients.}
    \label{fig:cmbnoisevar}
\end{figure}

\begin{figure}
    \includegraphics[width=\linewidth]{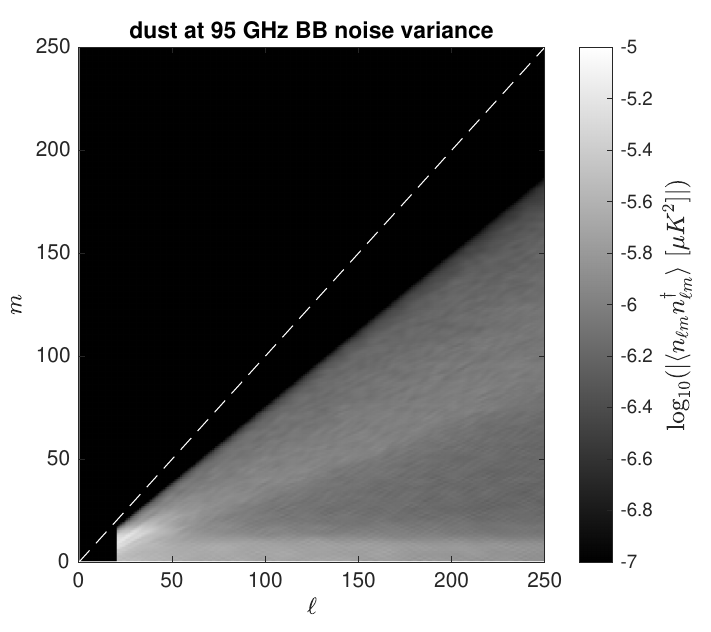}  
    \caption{Same as Fig.~\ref{fig:cmbnoisevar} but for the dust component map scaled to 95 GHz.}
    \label{fig:dustnoisevar}
\end{figure}

We multiply the $Q$ and $U$ maps with this weight in order to downweight noisy pixels and compute harmonic coefficients. Computing the noise variance per $(\ell,m)$ mode yields Figs.~\ref{fig:cmbnoisevar} and \ref{fig:dustnoisevar}. The noise variance for low-$m$ modes is significantly boosted in the CMB component, as they correspond to modes along the BICEP/\textit{Keck} scan direction which are taken out by filtering. The wedge patterns can be explained by the correspondence of constant-$m$ modes and the most negative observed declination: the higher the maximal declination, the higher the $m$ modes with non-zero power. As expected, the BICEP/\textit{Keck} noise is significantly lower than the \textit{Planck} noise when comparing low and high-$m$ modes in the CMB component. For the dust component, however, \textit{Keck} 220 GHz and \textit{Planck} 353 GHz contribute about equally to the sensitivity, and we obtain a more homogeneous noise variance.\\\\\\

\section{Maps}
In this section we will present simulated and real data CMB and dust component maps for BICEP, \textit{Keck}, and \textit{Planck} HFI data. For the reasons explained in the previous section, any plot of the map and computation of a power spectrum requires some weighting in the $(\ell,m)$ plane to avoid being dominated by large noise modes at low-$m$. \\

One option is to reobserve the CMB or dust component map again with one of the BICEP/\textit{Keck} observing matrices. This will produce a CMB or dust map as observed by one of the BICEP/\textit{Keck} receivers. For this, we choose the BICEP3 matrix since it has the largest footprint
$$
\mathbf{m}^\textrm{CMB/dust}=\mathbf{R}_\textrm{B3}\cdot\hat{\mathbf{s}}^\textrm{CMB/dust}
$$
This method also allows us to use the well-established pipeline from observed BICEP/\textit{Keck} maps, to purified power spectra, to estimates of $r$ \cite{bkmatrix,BK18}.\\ 

\begin{figure*}
    \centering
    \includegraphics[width=\linewidth]{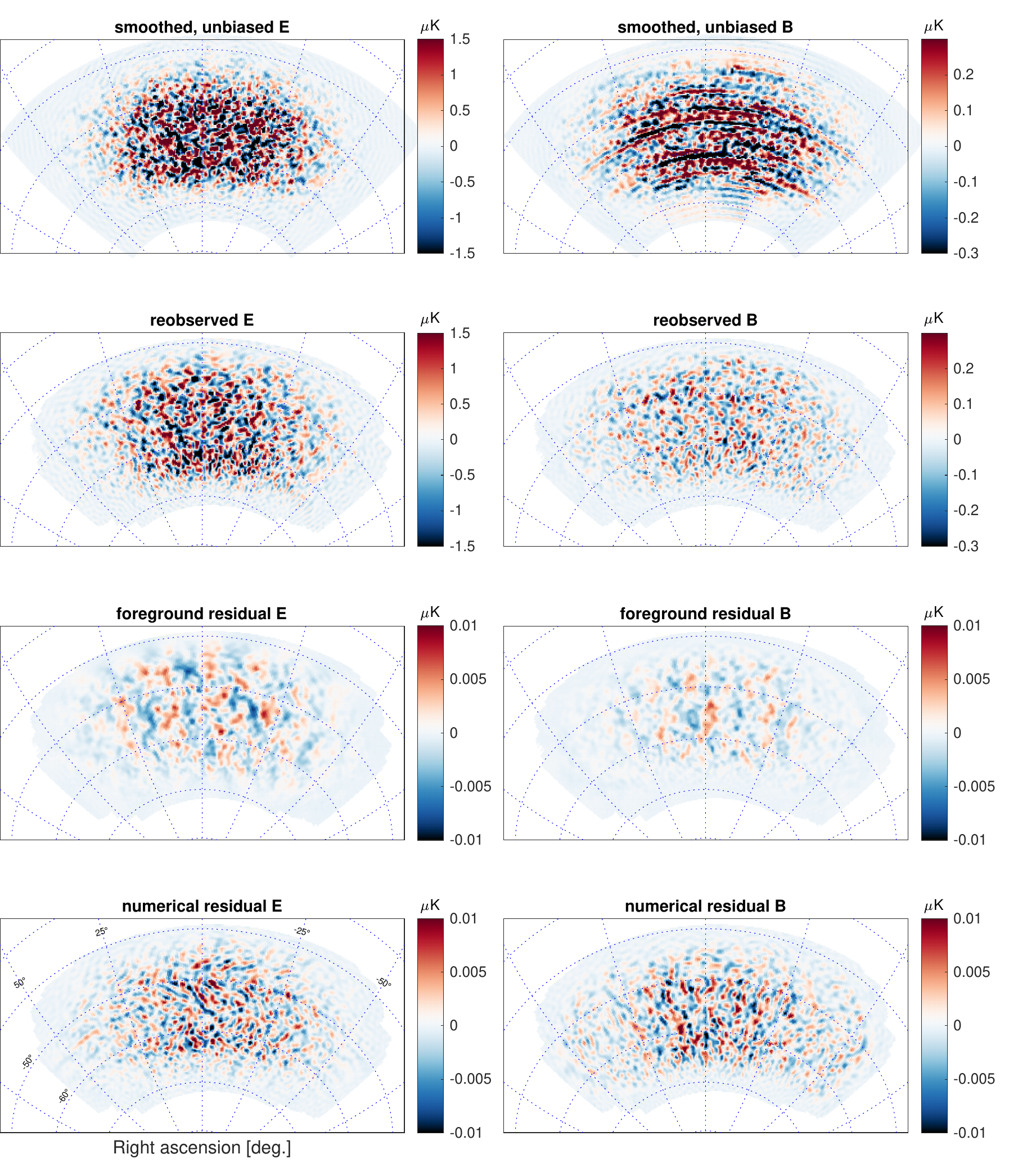}
    \caption{The CMB part of the component map estimator when run on a simulation, after apodization and $B$-mode purification.\\
    \textit{First row:} The input is a simulated realization of lensed $\Lambda$CDM,
    galactic dust and noise.
    The output has been convolved with a $20$ arcmin Gaussian beam, multiplied with a pixel-space weighting and transformed into $E$ and $B$-mode maps correcting for $E$-to-$B$-leakage effects from the masking \cite{Smith2005}.\\
    \textit{Second row:} After applying the BICEP3 observing matrix and purifying using the corresponding purification matrix \cite{bkmatrix}.\\
    \textit{Third row:} Subtracting the corresponding CMB-and-noise-only simulation from the maps in the previous row reveals the residual from foreground. Given that the foreground simulations used here are isotropic and Gaussian, the residual is entirely caused by statistical fluctuation of the $\beta_d$ fit.\\
    \textit{Fourth row:} To get an estimate of the numerical residual, we subtract the corresponding BICEP3 CMB-only simulation from the CMB component map simulation. This contains effects from the small differences in the filtering done on actual timestreams versus what is incorporated into the observing matrix, and numerical errors coming from the spherical harmonic transforms and the iterative solution method applied to obtain the CMB component map estimate. Note the much reduced color range in rows three and four.}
    \label{fig:ebsims}
\end{figure*}

In Fig.~\ref{fig:ebsims} we show the CMB part of the component map estimator when running on a simulated realization of lensed $\Lambda$CDM, Galactic dust and noise. The first row is the raw result, and the second row applies the BICEP3 observing and purification matrices. The observing matrix filters the noise dominated low-$m$ modes in the $B$-mode map, making the lensing $B$-modes visible. In the third row, by subtracting the corresponding CMB-and-noise-only simulation, we obtain the foreground residual map caused by the statistical fluctuation of the $\beta_d$ estimate. In the fourth row, we subtract a CMB-only single-frequency simulation from the corresponding CMB-only simulation run through the component separation pipeline to obtain the numerical residual introduced by the implementation of the component separation estimator. Both foreground and numerical residuals are well below the $E$- and $B$-mode signal and noise in the CMB component map. We will quantify their amplitude and impact on science results at the level of power spectra below.\\

\begin{figure*}
    \centering
    \includegraphics[width=\linewidth]{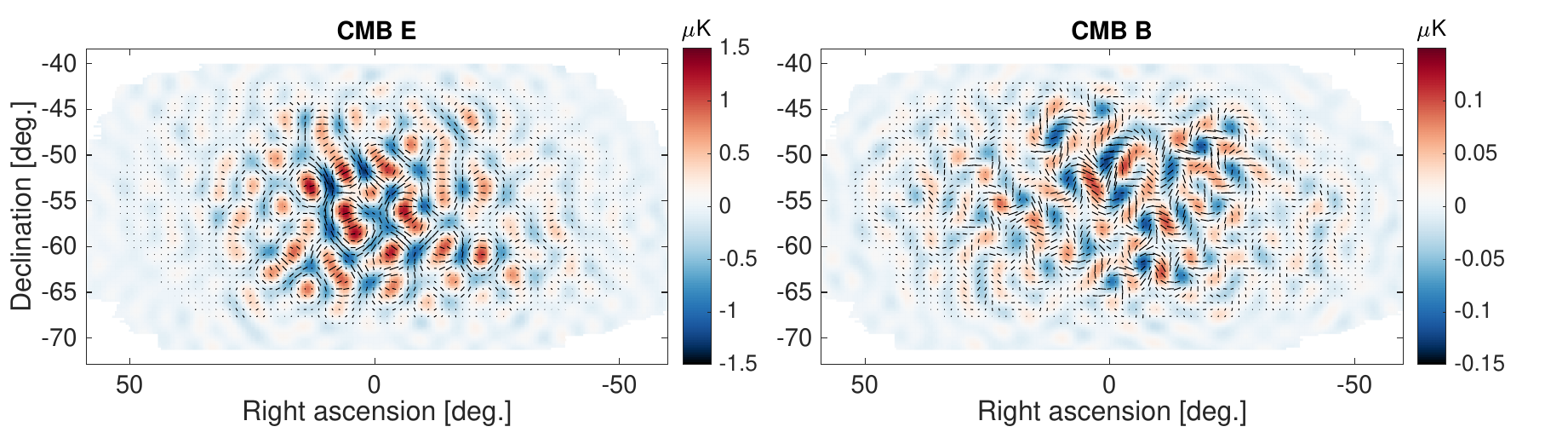}
    \caption{\emode\ (left) and \bmode\ (right) maximum-likelihood maps of the CMB in CMB units, beam-convolved and filtered like a \bicepthree\ map at 95 GHz with an additional bandpass filter to degree angular scales ($50<\ell<120$). Note the differing color ranges; on the left, the $E$ map is dominated by \lcdm\ signal, whereas on the right the $B$ map is approximately equal parts lensed-\lcdm\ signal and noise.}
    \label{fig:ebmaps_cmb}
\end{figure*}

\begin{figure*}
    \centering
    \includegraphics[width=\linewidth]{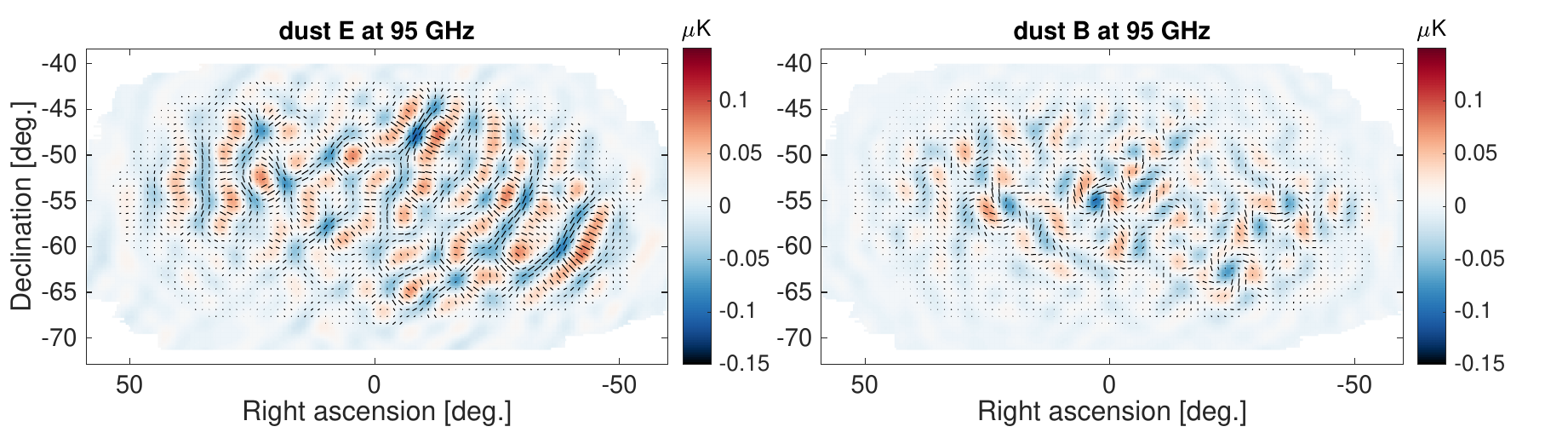}
    \caption{\emode\ (left) and \bmode\ (right) maximum-likelihood maps of thermal dust in CMB units, beam-convolved and filtered like a \bicepthree\ map at 95 GHz with an additional bandpass filter to degree angular scales ($50<\ell<120$). The maps are apodized and $B$-modes are purified. The $E$ modes are visibly brighter than the $B$ modes.}
    \label{fig:ebmaps_dust}
\end{figure*}

\begin{figure*}
    \centering
    \includegraphics[width=\linewidth]{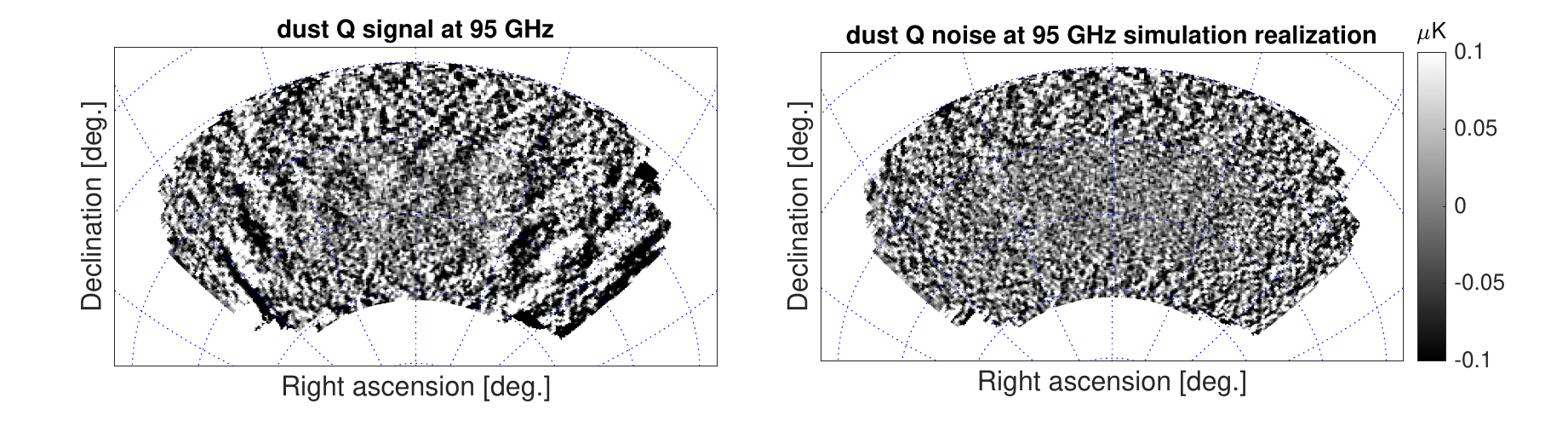}
    \includegraphics[width=\linewidth]{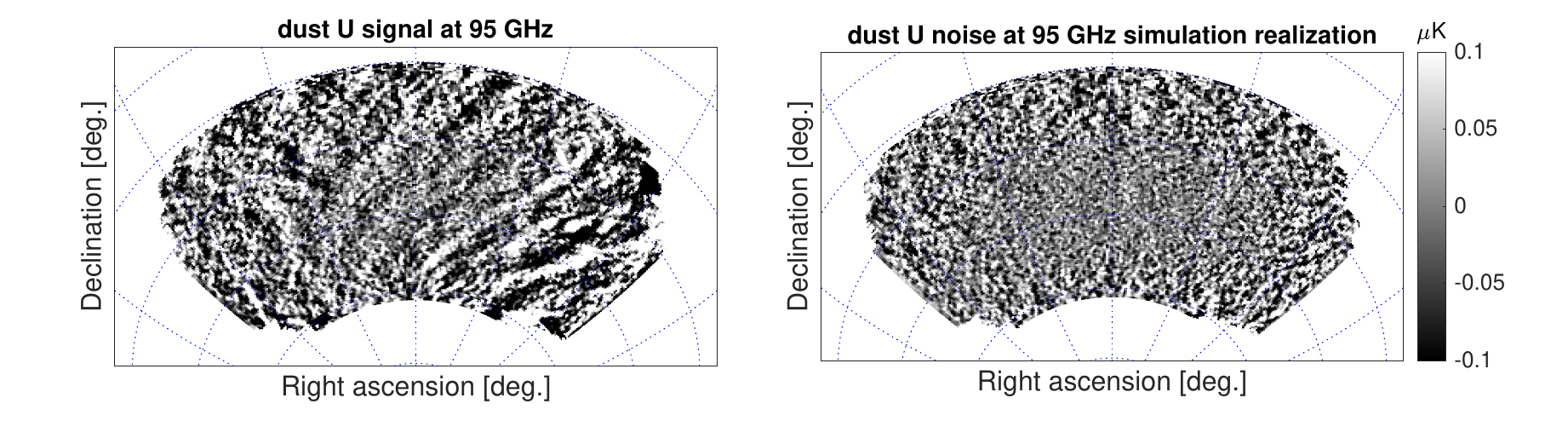}
    \includegraphics[width=\linewidth]{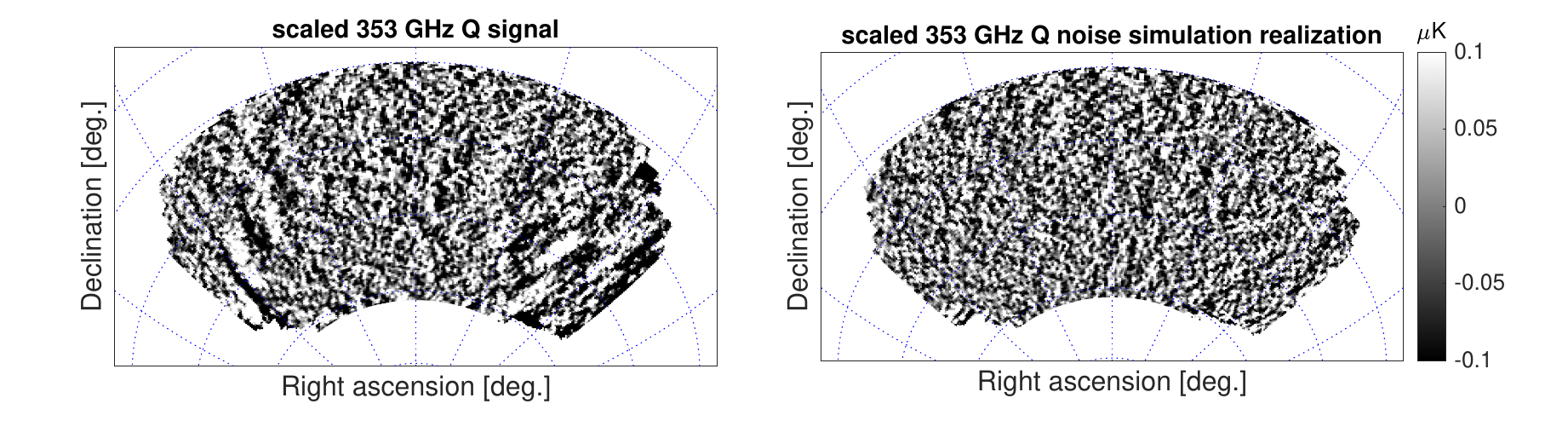}
    \includegraphics[width=\linewidth]{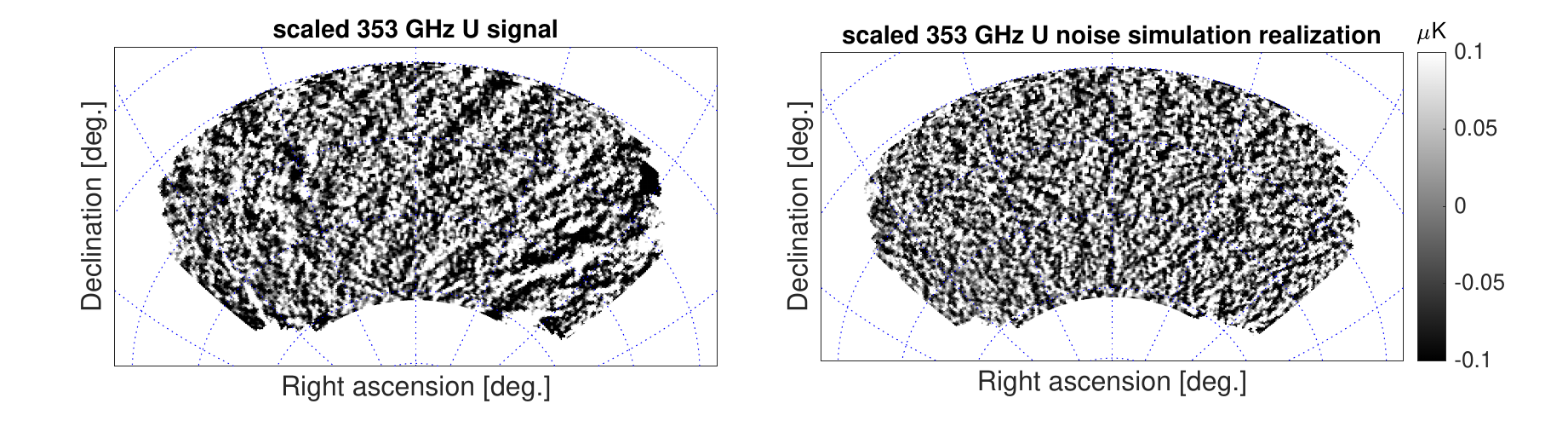}
    
    \caption{The first and second rows show the derived thermal dust component maps of this paper for $Q$ and $U$ signal (left), respectively, compared to a realization of noise (right). The noise is lower in the central BICEP2/\textit{Keck} region due to the high-frequency data from \textit{Keck}. Structure in the dust polarization is apparent, particularly in the outer regions of the observing field. The third and fourth rows show equivalent $Q$ and $U$ maps for \textit{Planck} 353\,GHz data (left) and one noise simulation (right), scaled to 95\,GHz using a modified blackbody (MBB) model with $\beta_d = 1.5$. These \textit{Planck} maps have been convolved to the same resolution as the derived component maps. All maps are reobserved with the \bicepthree\ observing matrix.}
    \label{fig:qumaps_dust}
\end{figure*}

The real data $E$ and $B$-mode maps of the CMB and the dust components are shown in Figs.~\ref{fig:ebmaps_cmb} and \ref{fig:ebmaps_dust}, respectively. Figure~\ref{fig:qumaps_dust} presents the derived dust component $Q$ and $U$ maps alongside a noise realization and a comparison with \textit{Planck} 353\,GHz data. The high-signal-to-noise filamentary dust structures are clearly visible in $Q$ and $U$, consistent with features previously identified in external neutral-hydrogen data \cite{bkhi} as well as SPT-3G \cite{sptbb}.\\\\

\section{Power Spectrum Estimation}

As described in the previous section, the inhomogeneous noise in the CMB and dust component maps requires some kind of weighting in harmonic space in order to downweight noisy modes. Considering this, we will investigate in this section how to optimally compute power spectra from these maps.

\subsection{Methods}

\subsubsection{Pseudo-$C_\ell$ method}

In order to use the pseudo-$C_\ell$ method, which is the standard method for estimating power spectra of BICEP/\textit{Keck} maps, we need to introduce some additional harmonic-space weighting to the standard pipeline. One natural way, as discussed in the previous section, is to apply a BICEP/\textit{Keck} observing matrix to the CMB and dust component maps. This not only applies a low-$m$ filter but also renders these maps into the right format to use the existing power-spectrum estimation pipeline based on computing pseudo-$C_\ell$s of matrix-purified maps as outlined in Ref.~\cite{bkmatrix}.

\subsubsection{Optimal quadratic maximum likelihood (QML) method}
Following Refs.~\cite{Tegmark1997,Bunn2017}, we can construct an optimal power spectrum estimator as the solution of a likelihood maximization by estimating a given bandpower at bin $b$ as
$$
\mathbf{D}^{BB}_b \sim \mathbf{m}^T \mathbf{C}^{-1} \mathbf{P}_b \mathbf{C}^{-1} \mathbf{m},
$$
where $\mathbf{m}$ are the input maps and the total covariance 
$$
\mathbf{C}\equiv\mathbf{S}+\bar{\mathbf{N}}
$$
is the sum of the signal and noise covariance. The operator $\mathbf{P}_b$ is given by
$$
\mathbf{P}_b=\frac{\partial \mathbf{C}}{\partial D^{BB}_b}.
$$
We model the signal covariance to be diagonal in harmonic space, given our baseline fiducial cosmological model derived from \textit{Planck} 2013 cosmological parameters \cite{planck2013XVI}
$$
\mathbf{S}=\mathbf{Y}^\dagger \textrm{diag}\left(\mathbf{C}_\ell^{EE},\mathbf{C}_\ell^{BB}\right) \mathbf{Y},
$$
where $\mathbf{Y}$ and $\mathbf{Y}^\dagger$ are forward and backward spherical harmonic transformations, respectively. The noise covariance is modeled using noise-only simulations as
$$
\bar{\mathbf{N}}^{-1}=\mathbf{w} \mathbf{Y}^\dagger \mathrm{diag}\left( N_{\ell m} \right)^{-1} \mathbf{Y} \mathbf{w},
$$
where $\mathbf{w}$ is the pixel weight in Fig.~\ref{fig:mapvar} and $N_{\ell m}$ is the noise variance per $(\ell,m)$-mode in Fig.~\ref{fig:cmbnoisevar}. In the actual implementation of the QML, we construct the inverse of the total covariance as
$$
\mathbf{C}^{-1}=\left(\bar{\mathbf{N}}^{-1}\mathbf{S}+\mathbf{1}\right)^{-1}\bar{\mathbf{N}}^{-1},
$$
such that we are never required to actually build $\bar{\mathbf{N}}$. The inverse is computed using singular value decomposition. In Ref.~\cite{Bunn2017}, a pure-B QML estimator was proposed by introducing a free parameter $\alpha$ in the signal covariance matrix
$$
\mathbf{S}=\mathbf{Y}^\dagger \textrm{diag}\left(\alpha\mathbf{C}_\ell^{EE},\mathbf{C}_\ell^{BB}\right) \mathbf{Y}.
$$
In the limit where $\alpha$ tends to infinity, we can write the inverse of this matrix as
$$
\mathbf{S}^{-1}=\mathbf{Y}^\dagger \textrm{diag}\left(0,1/\mathbf{C}_\ell^{BB}\right) \mathbf{Y}.
$$
With this inverse and the inverse of the noise covariance matrix, we can write the required inverse of the total covariance matrix as
$$
\mathbf{C}^{-1}=\mathbf{S}^{-1}\left(\bar{\mathbf{N}}^{-1}+\mathbf{S}^{-1}\right)^{-1}\bar{\mathbf{N}}^{-1}.
$$

\subsection{Simulation validation}
\begin{figure}
    \centering
    \includegraphics[width=\linewidth]{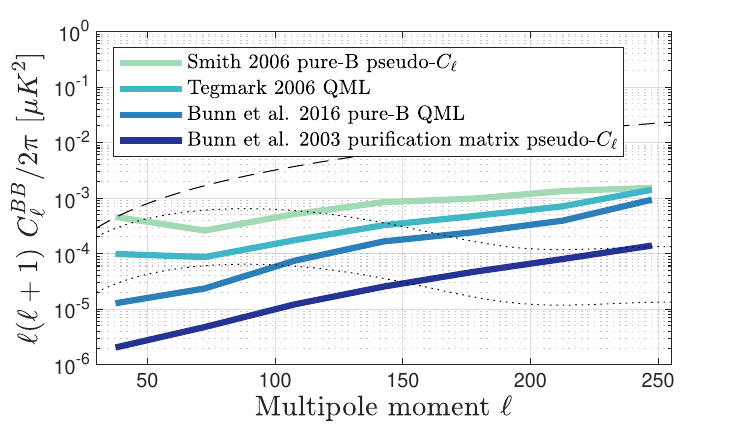}
    \caption{Comparing purification performance for different power spectrum estimators considered in this work: a simple pseudo-$C_\ell$ pure-B estimator of Ref.~\citep{Smith2005}, optimal quadratic maximum likelihood (QML) methods of Refs.~\citep{Tegmark1997} and \citep{Bunn2017}, and a purification-matrix-based method described in Ref.~\cite{Bunn2003} and used in previous BICEP/\textit{Keck} analyses \cite{bkmatrix}. The dashed line corresponds to the lensed-$B$ power of our baseline $\Lambda$CDM model, while the dotted lines correspond to a primordial gravitational wave signal of $r=10^{-2}$ and $r=10^{-3}$.}
    \label{fig:pureaps}
\end{figure}

We run the power spectrum estimators introduced in the previous section on the standard set of 499 BK18 simulations. We test for the level of $E$-to-$B$ leakage by estimating a $B$-mode auto-power spectrum on simulation maps which have no $B$-mode power in the input maps. In particular, we use unlensed $\Lambda$CDM simulations. The residual $B$-mode power spectra are shown in Fig.~\ref{fig:pureaps}.\\

We find that simply accounting for the mask-induced mode-coupling with the pure-$B$ estimator of Ref.~\cite{Smith2005,Grain2009} is insufficient for our sensitivity. $B$-mode purity can be improved with a QML estimator, in addition to the obvious advantage of optimal sensitivity. The purification-matrix-based method performs best in this purification test.\\

\begin{figure}
    \centering
    \includegraphics[width=\linewidth]{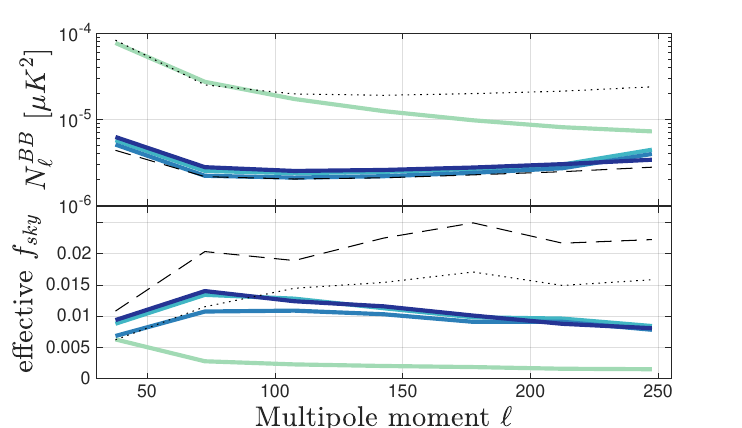}
    \caption{{\it First row:} The noise spectra for the different power-spectrum estimators run on the CMB component map. Colors are the same as in Fig.~\ref{fig:pureaps}. The BICEP3
    95\,GHz noise and the \textit{Keck} 220\,GHz noise are shown in the black dashed and dotted lines, respectively.
    The spectra are shown after correction for the filtering of signal which occurs
    due to the beam roll-off, timestream filtering, and \bmode\ purification.
    (Note that no $\ell^2$ scaling is applied.)
    {\it Second row:} The effective sky fraction as calculated from
    the ratio of the mean noise realization bandpowers to their fluctuation
    $f_\mathrm{sky}(\ell)=\frac{1}{2\ell \Delta \ell} \left( \frac{\sqrt{2}\bar{N_b}}{\sigma(N_b)} \right)^2$,
    i.e.\ the observed number of B-mode degrees of freedom divided by the
    nominal full-sky number.}
    \label{fig:fsky}
\end{figure}

We show the sensitivity to the CMB component for each estimator in Fig.~\ref{fig:fsky} in terms of the noise power $N_\ell$ and effective degrees of freedom. For the simple pseudo-$C_\ell$ estimator, the noise weighting is highly sub-optimal. The other estimators perform very similarly. The noise is only a little elevated compared to the noise in the BICEP3 95\,GHz frequency map. For low multipoles, the effective $f_{sky}$ is around $1-1.5\%$ and hence lies between the effective sky fraction of the larger BICEP3 and the smaller \textit{Keck} footprints, as expected. This motivates us to select the matrix-based purification method as the baseline for the remainder of this paper, as this method allows us to cut ambiguous modes quite aggressively without sacrificing a noticeable amount of sensitivity.\\\\

\begin{figure}
    \centering
    \includegraphics[width=\linewidth]{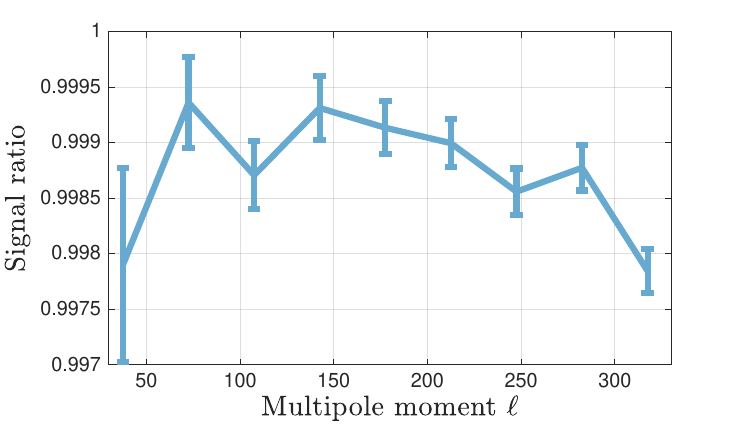}
    \caption{Suppression factor of the CMB component map computed as the ratio between the mean of power spectra of 499 signal-only simulations and the expectation for each bandpower. Error bars show the standard error of this mean.}
    \label{fig:signalratio}
\end{figure}

The CMB component map is supposed to be unbiased, meaning the suppression from timestream filtering and deprojection should be corrected for at the map-level. This is different compared to the baseline approach in the BICEP/\textit{Keck} analysis, where filtering is corrected for at the power-spectrum level. In Fig.~\ref{fig:signalratio} we show the residual suppression factor by plotting the ratio between the mean of the output power spectra for signal-only simulations and the expectation in each bandpower (which is likewise computed from signal simulations that are not run through the component separation estimator). We find a negligible, sub-percent residual signal suppression caused by the component separation estimator.\\

\begin{figure}
    \centering
    \includegraphics[width=\linewidth]{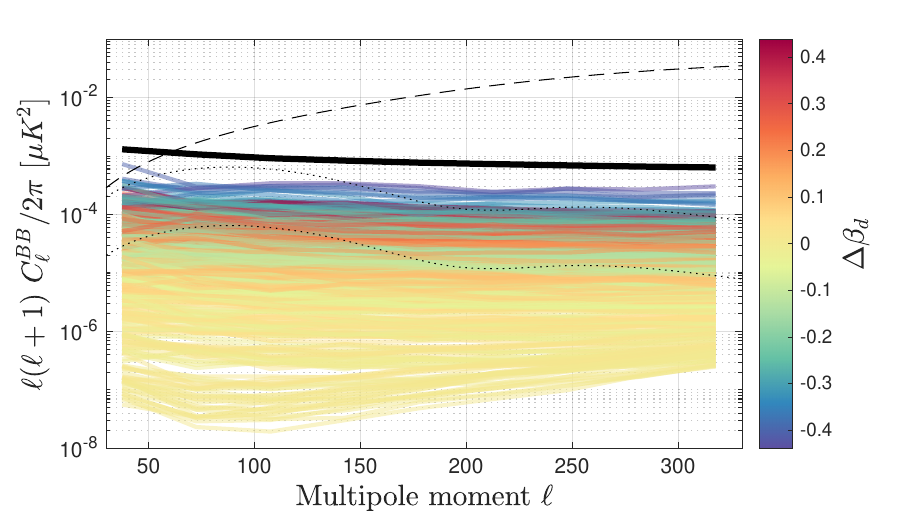}
    \caption{Foreground residual in the CMB component B-mode auto-power spectrum for 499 Gaussian-dust-only simulations. The color of the lines show the deviation of the respective best-fit value of $\beta_d$ for the specific realization from the simulation input of $\beta_d^\textrm{true}=1.6$. The thick solid black line indicates the dust level at 95 GHz, without any foreground cleaning, as measured in Ref.~\cite{BK18} The dashed line corresponds to the lensed-$B$ power of our baseline $\Lambda$CDM model while the dotted lines correspond to a primordial gravitational wave signal of $r=10^{-2}$ and $r=10^{-3}$.}
    \label{fig:dustresidual}
\end{figure}

We further can estimate the residual dust power in the CMB component map. We distinguish between statistical and systematic residual \cite{Stompor2016}. The former is caused by the statistical uncertainty of the dust model used to build the estimator. Specifically, the estimated $\beta_d$ scatters around the true simulation-input value of $\beta_d^\textrm{true}=1.6$. Propagating this scatter to the CMB $B$-mode power spectrum using dust-only simulations as input results in the statistical residuals shown in Fig.~\ref{fig:dustresidual}. This statistical residual will be included in the error budget of the CMB B-mode bandpowers and covariance.\\

\begin{figure}
    \centering
    \includegraphics[width=\linewidth]{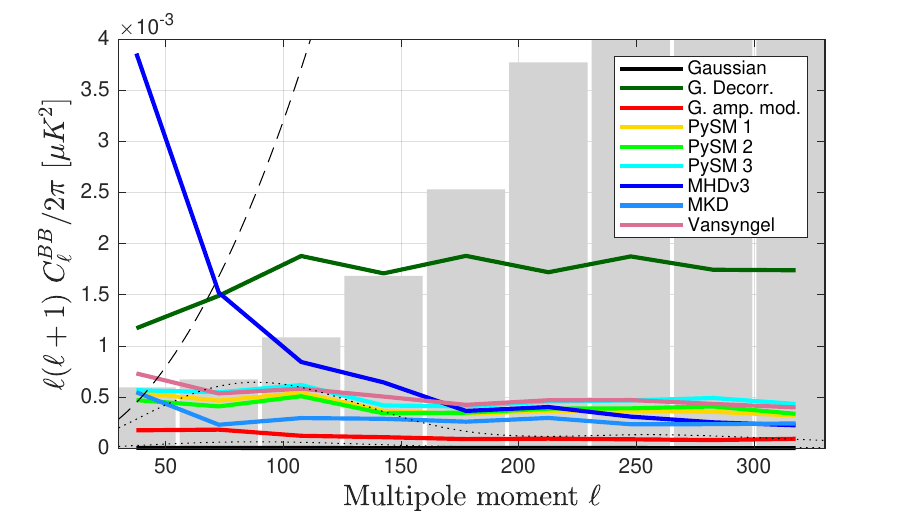}
    \caption{Systematic bias from one realization of the Gaussian and the alternate dust models as described in Refs.~\cite{biceptwoX,BK18} and Sec.~\ref{sec:dustmodels}. The gray bars represent the standard deviation of $\Lambda$CDM+dust+noise simulations in each bandpower. Most models lead to residuals smaller than this scatter representing cosmic variance. The dashed line corresponds to the lensed-$B$ power of our baseline $\Lambda$CDM model.}
    \label{fig:dustresidual_altdust}
\end{figure}

\label{sec:dustmodels}
We estimate the potential systematic residual, i.e.\ the residual caused by any mismatch between the true dust model and the model assumed in the estimator, by running the component separation on our suite of more complex alternate dust models \cite{BK18}. In Fig.~\ref{fig:dustresidual_altdust} we show the systematic residuals for all eight alternate dust models described in Refs.~\cite{biceptwoX,BK18}. The PySM models 1–3 use \textit{Planck}- and WMAP-derived templates for dust and synchrotron with varying spectral models and added small-scale Gaussian structure, though they overpredict dust levels in the BICEP/\textit{Keck} field \cite{Thorne2017,panex2025}. The MHDv2 model simulates non-Gaussian dust and synchrotron emission from a 3D Galactic magnetic field \cite{Kritsuk2018}. The MKD model adds 3D dust structure with varying density, temperature, and spectral index \cite{MartinezSolaeche2018}, while the Vansyngel model constructs Q/U maps by integrating over multiple magnetic field layers with fixed intensity and varying polarization \cite{Vansyngel2018}.\\

The Gaussian-decorrelation and MHDv3 models lead to significantly large residuals. The former incorporates a decorrelation parameter of $\Delta_d=0.85$ following the parametrization presented in Ref.~\cite{biceptwoX}, which is much lower than what is allowed by data in the BICEP/\textit{Keck} patch \cite{BK18}. Hence such a residual is expected and could be accounted for by marginalizing over a decorrelation parameter in the baseline BICEP/\textit{Keck} analysis \cite{Azzoni2023,BK18}. The MHDv3 model contains a significant amount of polarized synchrotron emission, about a factor four larger than the 95\% C.L. upper limit reported in Ref.~\cite{BK18}. Given that we do not presently model a synchrotron component in the map-based component separation, such a residual is expected. All other alternate foreground models, most notably the PySM and Vansyngel models incorporating spatial variation of the dust SED, are well below the 1$\sigma$ error bars in each bandpower and comparable to the statistical dust residual.\\

\begin{figure}
    \centering
    \includegraphics[width=\linewidth]{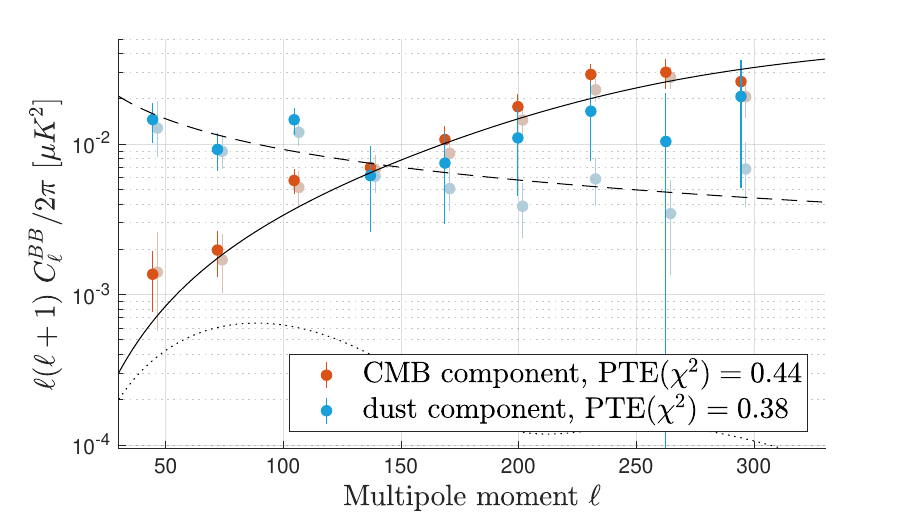}
    \caption{Real-data bandpowers of the CMB (orange points) and dust (blue points) component maps. The solid black line is the fiducial $\Lambda$CDM $B$-mode power spectrum from CMB lensing, the dotted black line is the primordial $B$-mode power spectrum for $r=10^{-2}$, and the dashed black line corresponds to the best-fit dust model of Ref.~\cite{BK18} at 150\,GHz. The faint points are derived using the cross-frequency likelihood method, where results represent a full MCMC sampling of the posterior within each individual bandpower bin; these are equivalent to Fig.~16 of BK18~\cite{BK18} but with marginalization over synchrotron parameters removed. The $\chi^2$ PTE values for the CMB and dust components, comparing the data bandpowers to the $\Lambda$CDM and best-fit dust models, are 0.44 and 0.38, respectively.}
    \label{fig:realaps}
\end{figure}

Finally, we show in Fig.~\ref{fig:realaps} the bandpowers of the real data CMB and dust component maps together with the theory power spectrum of $\Lambda$CDM and the best-fit dust model of BK18 \cite{BK18}. Using 499 simulations, we compute the probability-to-exceed (PTE) of the $\chi^2$ values between the data bandpowers and the best-fit model, obtaining PTEs of 0.44 for the CMB component and 0.38 for the dust component. We therefore conclude that these bandpowers are consistent with the best-fit $\Lambda$CDM and dust models, respectively. Notably, this implies that the dust bandpowers are consistent with a power-law shaped spectrum.\\

\section{Consistency with Multi-Frequency-Spectra Likelihood}

The component separation method presented in this work allows us to build an alternative pipeline for estimating the tensor-to-scalar ratio $r$. We check for consistency between the two pipelines at the level of estimated parameters. In the baseline analysis, we estimate $r$ together with seven foreground parameters using multi-frequency power spectra computed from BICEP/\textit{Keck}, WMAP, and \textit{Planck} frequency maps \cite{biceptwoX,BK18}. In this work, we use a subset of these maps, in particular, the BICEP/\textit{Keck} and \textit{Planck} HFI maps, to compute CMB and dust component maps, from which we can compute CMB and dust B-mode auto-power spectra, as well as the cross-spectrum between the two.\\

We use the parametric likelihood function described in Ref.~\cite{BKP} for the two auto-power spectra and one cross-power spectrum computed from the CMB and the dust map. The bandpower covariance matrix is derived from
499 simulations of signal and noise, explicitly setting covariances between the CMB and dust signal-only bandpowers to zero, but allowing for noise covariances between the two components. We fit a one-parameter model consisting of a primordial tensor component with varying amplitude $r$ and a fixed component from CMB lensing, derived from the \textit{Planck} best-fit $\Lambda$CDM cosmology \cite{Planckparameters}, to the CMB auto-power spectrum
\begin{equation}
    {D}^{BB}_\ell=r\cdot D_\ell^{BB\ \mathrm{tensor}} + D_\ell^{BB\ \mathrm{lensing}}.
\end{equation}
The tensor-to-scalar ratio $r$ is evaluated at a pivot scale of $0.05\ \mathrm{Mpc}^{-1}$. At the same time, we fit a power-law model to the dust auto-power spectrum, referenced at a frequency $\nu=353\ \textrm{GHz}$,
\begin{equation}
    {D}^{BB}_\ell=A_\textrm{d} \cdot \left( \frac{\ell}{80} \right)^{\alpha_\textrm{d}}.
\end{equation}
\\
 
\begin{figure}
    \centering
    \includegraphics[width=\linewidth]{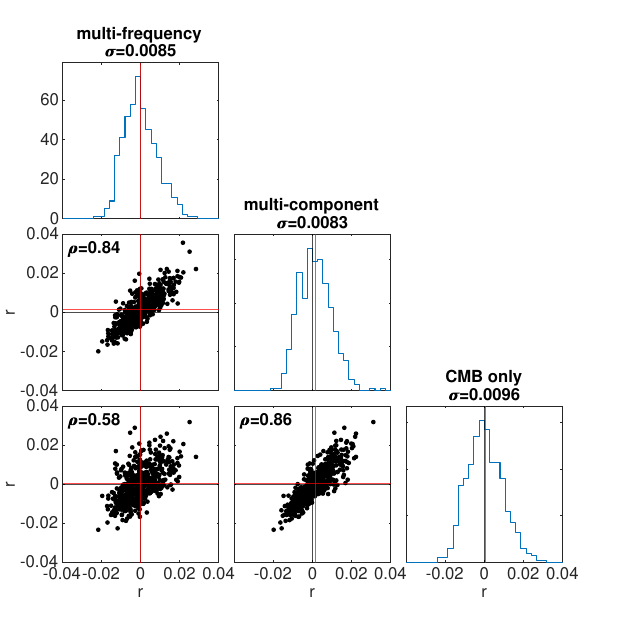}
    \caption{Histograms and 2D scatter plots between best-fit $r$ values obtained with the multi-frequency-spectrum-based
    method of Ref.~\cite{BK18} (with the modification of fixing the synchrotron component in the model and aligning the selected frequency bands with those used in this paper) and the multi-component-spectrum-based method presented in this paper, either including or excluding the dust component map in the power spectrum computation. Black vertical and horizontal lines show the simulation input of $r=0$, while the corresponding red lines show the distribution mean. The distribution standard deviations are shown in the respective histogram's title. The Pearson correlation coefficient, $\rho$, ranges from 58\% to 86\%.}
    \label{fig:mlcorr}
\end{figure}

In Fig.~\ref{fig:mlcorr} we show the distribution of best-fit values of $r$, computed for our set of 499 signal and noise simulations, for three likelihood variations:
\begin{itemize}
    \item ``multi-frequency": The baseline likelihood of the BK18 analysis \cite{BK18}, modified by fixing all synchrotron parameters in the model to their fiducial values and aligning the frequency bands with those used in this paper, to enable a direct comparison with the map-based approach presented here,
    \item ``multi-component": The likelihood introduced in the paragraph above, based on computing auto- and cross-power spectra between the CMB and dust component maps,
    \item ``CMB only": The same likelihood as above, but using only the CMB auto-power spectra.
\end{itemize}
The histograms show an unbiased recovery of the tensor-to-scalar ratio, with comparable sensitivity between the multi-frequency and multi-component likelihood approaches. Excluding any information about the dust power spectrum, and thus rendering it agnostic to the particular spectral shape of dust, degrades the sensitivity by about 16\%.\\

The Pearson correlation coefficient between these three distributions is $84\%$ for the multi-frequency and multi-component approaches. In \cite{s4compsep}, the correlation between the two methods was found to be higher, indicating that the departure from unit correlation in our analysis arises from the filtering applied to the real data. In the multi-frequency approach, each frequency map is filtered differently, whereas in the multi-component approach we attempt to undo this filtering, supplementing the missing modes with information from \textit{Planck}. This results in effective weighting differences, both in pixel and harmonic space, between the two methods. They are ultimately sensitive to slightly different modes. When the dust-component channel is removed from the likelihood, the correlation declines further to $56\%$.\\

Given the results presented in this section, we expect the map-based (multi-component) approach to deliver constraints on the tensor-to-scalar ratio that are consistent with those obtained using the baseline multi-frequency method. For the latest and most stringent upper-limit constraints on $r$, we therefore refer the reader to BK18\,\cite{BK18}.\\

\section{Conclusions}
We present, for the first time, component-separated maps derived from BICEP and \textit{Keck} data. Constructing such maps from frequency data with differing beams, filtering, and sky coverage requires accounting for mode coupling at the map level. We develop and validate a method to recover unbiased CMB and Galactic dust component maps, demonstrating optimal procedures to extract power spectra and cosmological parameters from these maps while carefully treating their non-trivial noise properties. Applying this approach to real data, we present the resulting CMB and dust power spectra and compare constraints on the tensor-to-scalar ratio, $r$, to those obtained with the baseline multi-frequency power-spectrum method of Ref.~\cite{BK18}. We find an $84\%$ correlation between recovered $r$ values, with a comparable $\sigma(r)$ from 499 simulations.\\

We find that including the dust component in the likelihood is essential for the map-level cleaning approach to achieve $r$ sensitivity comparable to the baseline BK18\,\cite{BK18} method. Omitting the dust channel instead provides a constraint that is agnostic to the assumed dust power-spectrum shape. A further strength of the map-based method lies in the production of high-fidelity CMB and dust component maps, which reveal the BICEP2/\textit{Keck} field with striking visual clarity, both in the clean CMB fluctuations and in the detailed Galactic dust structure, and enable analyses beyond simple power-spectrum estimation, such as higher-order statistics and cross-correlation studies. The dust component map produced here also serves as a high-quality foreground template, which has already proven valuable for assessing the impact of Galactic dust on cosmic birefringence measurements with BICEP3 \cite{bkbire}.\\

\begin{acknowledgements}

The \bicep/\keck\ projects have been made possible through a series of grants from the National Science Foundation most recently including 2220444-2220448, 2216223, 1836010, and 1726917. The development of antenna-coupled detector technology was supported by the JPL Research and Technology Development Fund and by NASA Grants 06-ARPA206-0040, 10-SAT10-0017, 12-SAT12-0031, 14-SAT14-0009, and 16-SAT-16-0002. The development and testing of focal planes was supported by the Gordon and Betty Moore Foundation at Caltech. Readout electronics were supported by a Canada Foundation for Innovation grant to UBC. Support for quasi-optical filtering was provided by UK STFC grant ST/N000706/1. The computations in this paper were run on the Odyssey/Cannon cluster supported by the FAS Science Division Research Computing Group at Harvard University. The analysis effort at Stanford and SLAC is partially supported by the U.S. DOE Office of Science.\\

We thank the staff of the U.S. Antarctic Program and in particular the South Pole Station without whose help this research would not have been possible. Most special thanks go to our heroic winter-overs during the observing seasons up until 2018: Robert Schwarz, Steffen Richter, Sam Harrison, Grantland Hall and Hans Boenish. We thank all those who have contributed past efforts to the \bicep/\keck\ series of experiments, including the \bicepone\ team. We also thank the \planck\ and \wmap\ teams for the use of their data.\\

We are grateful to Federico Bianchini and Radek Stompor for insightful discussions and to Josquin Errard for valuable comments on the manuscript.

\end{acknowledgements}

\bibliography{main}

\end{document}